\documentclass[iop,twocolappendix,numberedappendix,appendixfloats]{emulateapj}

\usepackage{natbib}
\usepackage[bookmarks,breaklinks]{hyperref}
   \hypersetup{
     colorlinks,
     citecolor=blue,
     filecolor=magenta,
     linkcolor=blue,
     urlcolor=cyan
}
\usepackage{multirow}
\usepackage{longtable}
\usepackage{threeparttablex}
\usepackage{csquotes}

\newcommand{\OIII}{\mbox{[O\,\textsc{iii}]}}
\newcommand{\NII}{\mbox{[N\,\textsc{ii}]}}

\newcommand{\Ha}{H$\alpha$}
\newcommand{\kms}{km s$^{-1}$}

\shorttitle{Comparing AGNs with and without strong outflows}
\shortauthors{Rongxin Luo et al.}

\begin{document}
\title{Unraveling the complex structure of AGN-driven outflows:\\ IV. Comparing AGNs with and without strong outflows}

\author{Rongxin Luo, Jong-Hak Woo, Jaejin Shin, Daeun Kang, Hyun-Jin Bae, Marios Karouzos}
\affil{Astronomy Program, Department of Physics and Astronomy, Seoul National University, Seoul 151-742, Republic of Korea}
\email{email: woo@astro.snu.ac.kr}

\begin{abstract}
AGN-driven outflows are considered as one of the processes driving the co-evolution of supermassive black holes with their host galaxies. 
We present integral field spectroscopy of six Type 2 AGNs at z $<$ 0.1, which are selected as AGNs without strong outflows based on the 
kinematics of \mbox{[O\,\textsc{iii}]} gas. Using spatially resolved data, we investigate the ionized gas kinematics and photoionization properties 
in comparison with AGNs with strong outflows. We find significant difference between the kinematics of ionized gas and stars for two AGNs, 
which indicates the presence of AGN-driven outflows. Nevertheless, the low velocity and velocity dispersion of ionized gas indicate relatively 
weak outflows in these AGNs. Our results highlight the importance of spatially-resolved observation in investigating gas kinematics and identifying 
the signatures of AGN-driven outflows. While it is unclear what determines the occurrence of outflows, we discuss the conditions and detectability 
of AGN-driven outflows based on a larger sample of AGNs with and without outflows, suggesting the importance of gas content in the host galaxies.
\end{abstract}


\section{Introduction}
\label{sec:intro}
Since the discovery of the correlation between masses of super-massive black holes (SMBHs) and global properties 
of their host galaxies (\citealt{Ferrarese2000,Gebhardt2000}), the co-evolution of galaxies and SMBHs has 
become an important topic in studying galaxy formation and evolution (\citealt{Alexander2012,Kormendy2013,Heckman2014}). 
As AGN feedback may play a crucial role in regulating the growth of SMBHs and galaxies, various feedback mechanisms have 
been included in current semi-analytic models and numerical simulations to reproduce the properties of massive galaxies 
(e.g., \citealt{DiMatteo2005,Springel2005,Croton2006,Hopkins2006}), as well as the observed black hole mass correlations with 
host galaxy properties (\citealt{Kormendy2013}). 

As a potential channel of AGN feedback, outflows have been 
investigated in both local and distant AGN host galaxies (see \citealt{Elvis2000,Veilleux2005,Fabian2012} and \citealt{Heckman2014} 
for reviews). Based on different tracers of gas kinematics, a growing body of statistical works have revealed the evidence 
of non-gravitational motions in the narrow line regions (NLRs), indicating that gaseous outflows are prevalent among AGNs 
(e.g., \citealt{Nesvadba2008,Wang2011,Zhang2011,Harrison2012,Mullaney2013,Bae2014,Genzel2014,Woo2016,Wang2018, Rakshit2018}). 
Recent spatially resolved observations have began to map the detailed properties of AGN-driven outflows in a multi-phase 
view. Based on the integral-field-spectroscopy (IFS) observations of optical forbidden lines (e.g. \mbox{[O\,\textsc{iii}]} 
$\lambda$5007\AA\ line), extensive studies have measured the geometry, kinematics, and energy of ionized gas outflows in 
the NLRs of local AGNs and high-z QSOs (e.g., \citealt{Sharp2010,Storchi-Bergmann2010,Liu2013,Liu2013a,Rupke2013,McElroy2015,Karouzos2016,Karouzos2016a,Bae2017,Kang2018}).
With the advent of near-infrared IFS and radio interferometry observations, massive outflows of neutral and molecular gas 
have also been discovered and characterized in different objects and samples (e.g. \citealt{Feruglio2010,Cicone2012,Maiolino2012}). 
However, the role of AGN-driven outflows is not fully understood as the observational studies provide no strong constraint on 
how AGN-driven outflows quench or enhance star formation.

Using a large sample of Type 2 AGNs at low-z, \citet{Woo2016} performed a statistical study to constrain the properties and fraction 
of ionized gas outflows as well as their relation to AGN energetics (see also \citealt{Bae2014,Woo2017,Kang2017}). They find 
that ionized gas outflows are ubiquitous among luminous Type 2 AGNs.
In a series of studies for investigating the detailed properties of ionized gas outflows, \citet{Karouzos2016,Karouzos2016a}, \citet{Bae2017} 
and \citet{Kang2018} have carried out IFS observations of a luminosity-limited sample selected from \citet{Woo2016}. They have established 
a set of robust methods to properly perform spatial and kinematic decomposition of the ionized gas emission, with which they 
effectively identify AGN-driven outflows and constrain the size, velocity, and kinetic energy of outflows (see \ref{sec:analysis} for details).

In this paper, we present a spatially resolved study of six Type 2 AGNs,
which are identified as no/weak outflow AGNs by \citet{Woo2016}, due to the lack of strong signature of outflows in the SDSS spectra.
Using the Gemini/GMOS-IFU data, we investigate the differences of gas properties between AGNs with and without strong outflows. 
We describe the sample and observations in section \ref{sec:sample}, and data reduction and analysis in section \ref{sec:method}.
We present the main results in section \ref{sec:results}. Discussion and summary follow in section \ref{sec:discussion} and \ref{sec:summary}.

\section{Sample and observations}
\label{sec:sample}
\subsection{Sample selection}
\label{sec:selection}

\begin{figure}[tb]
\begin{center}
\includegraphics[width=0.48\textwidth,angle=0]{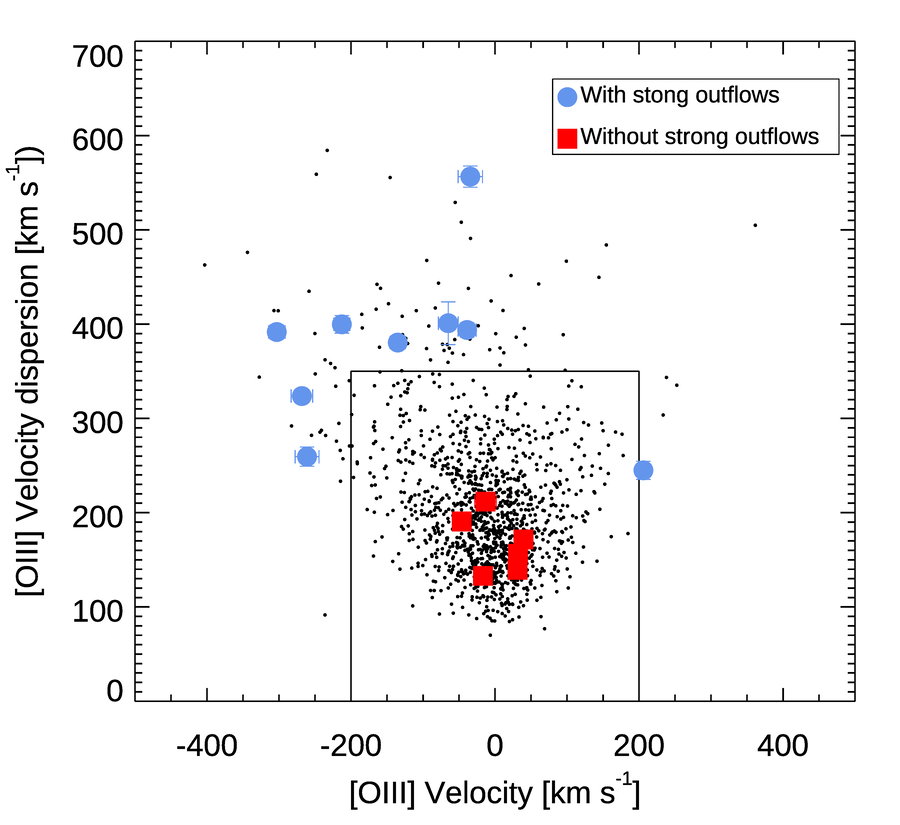}
\caption{\mbox{[O\,\textsc{iii}]} velocity versus velocity dispersion (VVD) diagram. The AGNs without strong outflows, which are selected for this study 
	(red circles) are compared to the AGNs with strong outflows in our previous studies (blue circles). The selection criteria of strong outflows are indicated 
	by the solid box as adopted by \citet{Karouzos2016, Kang2018}. The luminosity-limited sample of local AGNs is denoted with black dots \citep{Woo2016}.} 
\label{fig:sample}
\end{center}
\end{figure}

\begin{deluxetable*}{c c c c c c c c c c c c c}
	\tabletypesize{\footnotesize}
	\tablecolumns{10}
	\tablewidth{0pt}
	\tablecaption{Properties of the sample based on the SDSS spectra and the log of observations \label{tab:sample}}
	\tablehead{\colhead{ID}	&   \colhead{z}	&	\colhead{V$_{\mbox{[O\,\textsc{iii}]}}$}	&	\colhead{$\sigma_{\mbox{[O\,\textsc{iii}]}}$}	&	
	\colhead{$\log{\mbox{L}_{\mbox{[O\,\textsc{iii}]}}}$}	 & 	\colhead{$\log{\mbox{L}_{\mbox{[O\,\textsc{iii}];cor}}}$}	& \colhead{m$_{r}$}	& 
	\colhead{$\log{\mbox{M}_{\ast}}$} & \colhead{b/a} & \colhead{Date} & \colhead{t$_{\mathrm{exp}}$} & \colhead{Seeing}	& \colhead{AM}	\\
	\colhead{ }	& \colhead{ }	&	\multicolumn{2}{c}{[km s$^{-1}$]}	&	\multicolumn{2}{c}{[erg s$^{-1}$]}  &  \colhead{[AB]}	&
	\colhead{M$_{\odot}$} & 	\colhead{ }   &  \colhead{ }	&  \colhead{[min]}	&	\colhead{[\arcsec]}	&	\colhead{ }\\
	\colhead{(1)} & \colhead{(2)} & \colhead{(3)} & \colhead{(4)} & \colhead{(5)} & \colhead{(6)}  & \colhead{(7)} & \colhead{(8)} & \colhead{(9)} & \colhead{(10)}  & \colhead{(11)}  & \colhead{(12)} & \colhead{(13)}}
	\startdata

	J084344+354942  &	0.0541	&	-46	&   191	&	 41.7	& 42.7 & 14.70 & 11.07 &	    &	04/12/16	&	40.5	&	0.64	&	1.04\\
	J101936+193313  &	0.0647	&	-17	&   133	&	 41.4	& 42.1 & 16.20 & 10.51  &	0.60  &	04/12/16	&	96  	&	0.64	&	1.03\\
	J105833+461604  &	0.0397  &	 31	&   139	&	 40.9	& 41.5 & 13.94 & 11.06 &	0.89  &	04/12/16	&	40.5	&	0.64	&	1.14\\
	J115657+550821	&	0.0796	&	 32	&   156	&	 41.2	& 42.1 & 15.70 & 11.02 &	0.84  &	04/13/16	&	144	    &	1.00	&	1.22\\
	J131153+053138	&	0.0873	&	 39	&   172	&	 41.5	& 42.1 & 15.92 & 10.93  &	0.85  &	04/12/16	&	144	    &	0.64	&	1.04\\
	J161756+221943	&	0.1020	&	-13	&   212	&	 40.9	& 41.9 & 15.64 & 11.34 &	0.85  &	04/13/16	&	144	    &	1.00	&	1.04
	
	\enddata
	\tablecomments{
		Col. 1: target ID; Col. 2: redshift; Col. 3: \mbox{[O\,\textsc{iii}]} velocity shift with 
		respect to the systemic velocity; Col. 4: \mbox{[O\,\textsc{iii}]} velocity dispersion; Col. 5: dust-uncorrected \mbox{[O\,\textsc{iii}]} 
		luminosity; Col. 6: dust-corrected \mbox{[O\,\textsc{iii}]} luminosity (see \citealt{Bae2014}); Col. 7: \textit{r}-band magnitude from the 
		SDSS photometry; Col. 8: stellar mass from SDSS \citep{Bae2014, Woo2016}; Col. 9: minor-to-major axis ratio; Col. 10: date of GMOS 
		observations; Col. 11: exposure time; Col. 12: seeing; Col. 13: average airmass.}
\end{deluxetable*}

In order to compare the detailed properties of ionized gas in the AGNs with and without strong outflows, we selected 
AGNs with no strong signature of outflows from a large sample of $\sim$ 39,000 Type 2 AGNs at z $<$ 0.3, which was 
used in our previous statistical study (\citealt{Woo2016}) based on the archival spectra of the Sloan Digital Sky Survey. 
While AGNs with strong outflows were identified by tracing the extreme kinematic signatures of ionized gas manifested 
as large velocity shift or large velocity dispersion of \mbox{[O\,\textsc{iii}]}$\lambda$5007 emission line, we selected 
weak/no outflow AGNs using the following criteria. First, we limited the extinction-corrected \mbox{[O\,\textsc{iii}]} 
luminosity as $L_{\mathrm{\mbox{[O\,\textsc{iii}]};cor}} > 10^{42}$ erg s$^{-1}$ and set a redshift cut of z $< 0.1$, 
as applied for selecting strong outflow AGNs by \citet{Karouzos2016}. 
Second, we selected AGNs with small \mbox{[O\,\textsc{iii}]} velocity shift (i.e., $|v_{\mbox{[O\,\textsc{iii}]}}| < 50$ km s$^{-1}$) with respect 
to the systemic velocity, and low \mbox{[O\,\textsc{iii}]} velocity dispersion (i.e., \OIII\ velocity dispersion is consistent with stellar velocity 
dispersion within 20\%). The systemic velocity and stellar velocity dispersion ($\sigma_{\ast}$) were measured from stellar absorption lines 
in the SDSS spectra \citep{Woo2016}. In addition, we also matched stellar mass and the minor-to-major (b/a) axis ratio (i.e., inclination) of 
our targets with those of the AGNs with strong outflows in \citet{Karouzos2016}. As a final sample we chose six Type 2 AGNs for this study. 

The properties of the selected AGNs measured from the integrated SDSS spectra are presented in Table \ref{tab:sample}. 
In Fig. \ref{fig:sample}, we present the \mbox{[O\,\textsc{iii}]} velocity-velocity dispersion (VVD) diagram to contrast the gas kinematics 
between AGNs with and without strong outflows \citep[for details for AGNs with strong outflows, see][]{Karouzos2016,Kang2018},
along with the luminosity-limited sample of 902 AGNs (i.e., $L_{\mathrm{\mbox{[O\,\textsc{iii}]};cor}} > 10^{42}$ erg s$^{-1}$) at z $<$ 0.1
selected from \citet{Woo2016}. 

\begin{figure*}[tbp]
	\begin{center}
		\includegraphics[width=0.45\textwidth,angle=0]{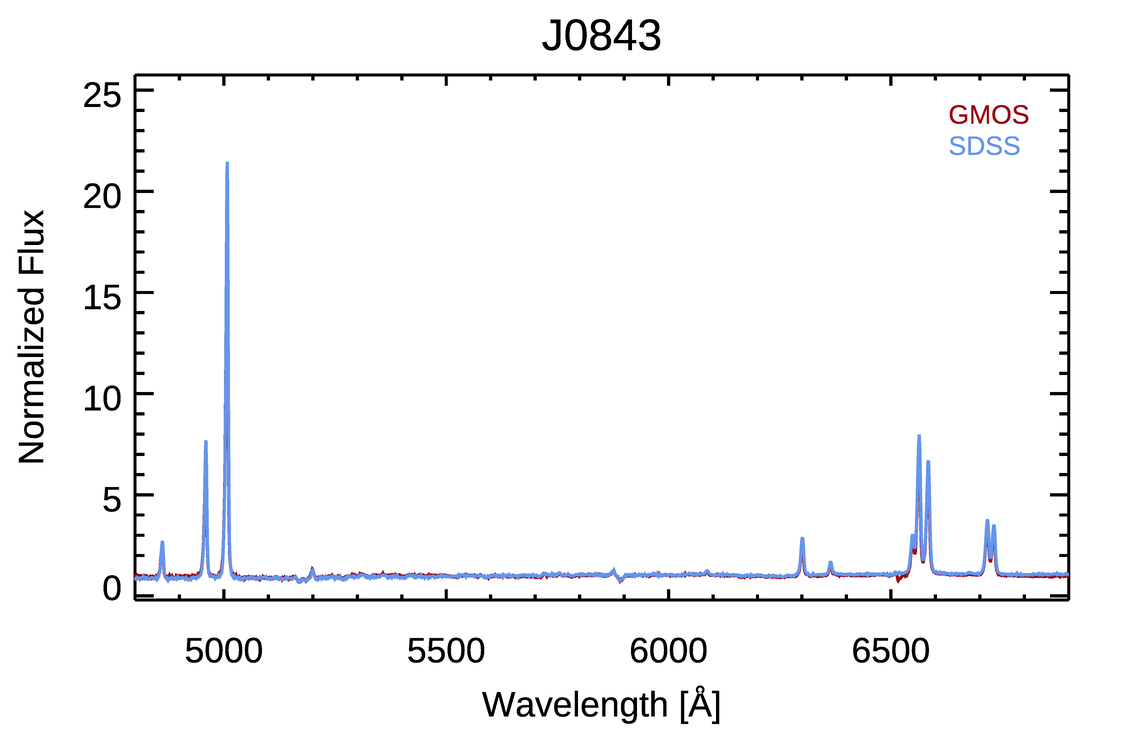}
		\includegraphics[width=0.45\textwidth,angle=0]{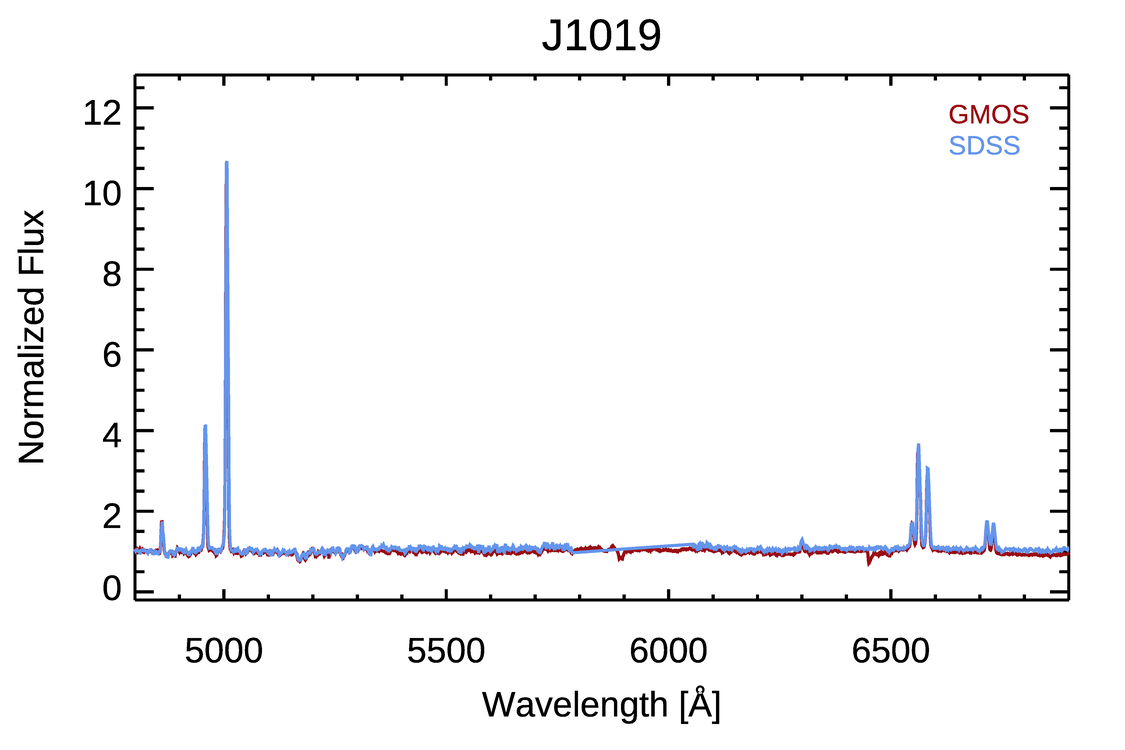}\\
		\includegraphics[width=0.45\textwidth,angle=0]{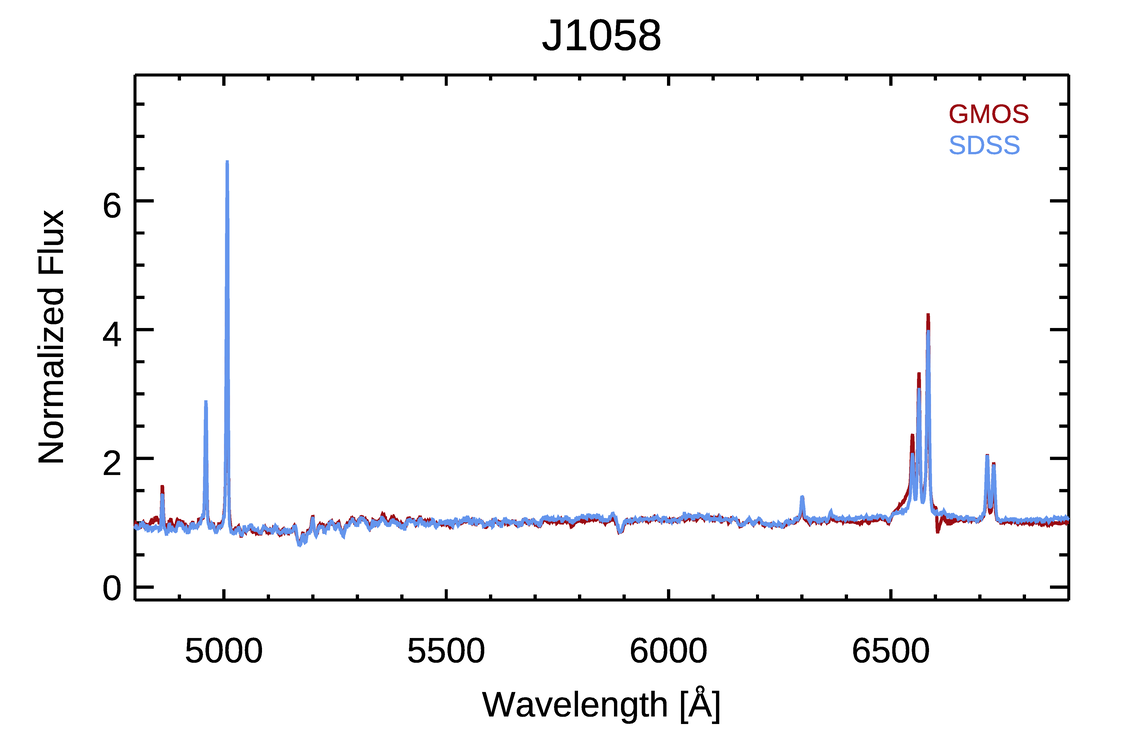}
		\includegraphics[width=0.45\textwidth,angle=0]{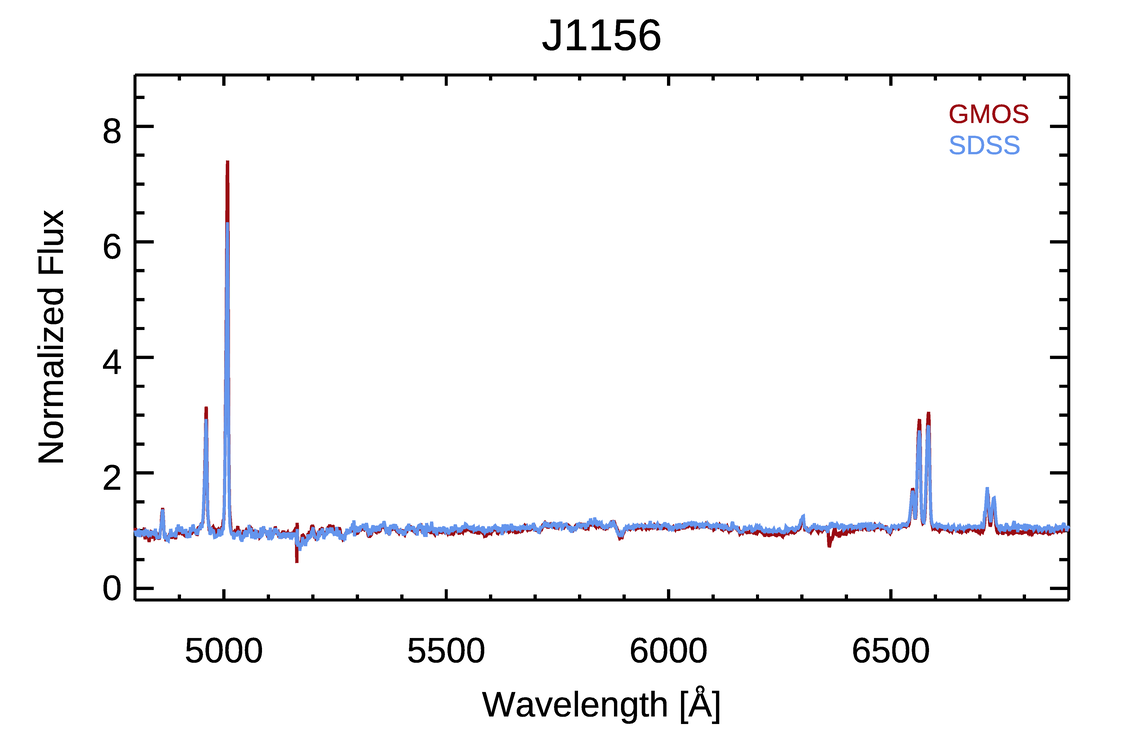}\\
		\includegraphics[width=0.45\textwidth,angle=0]{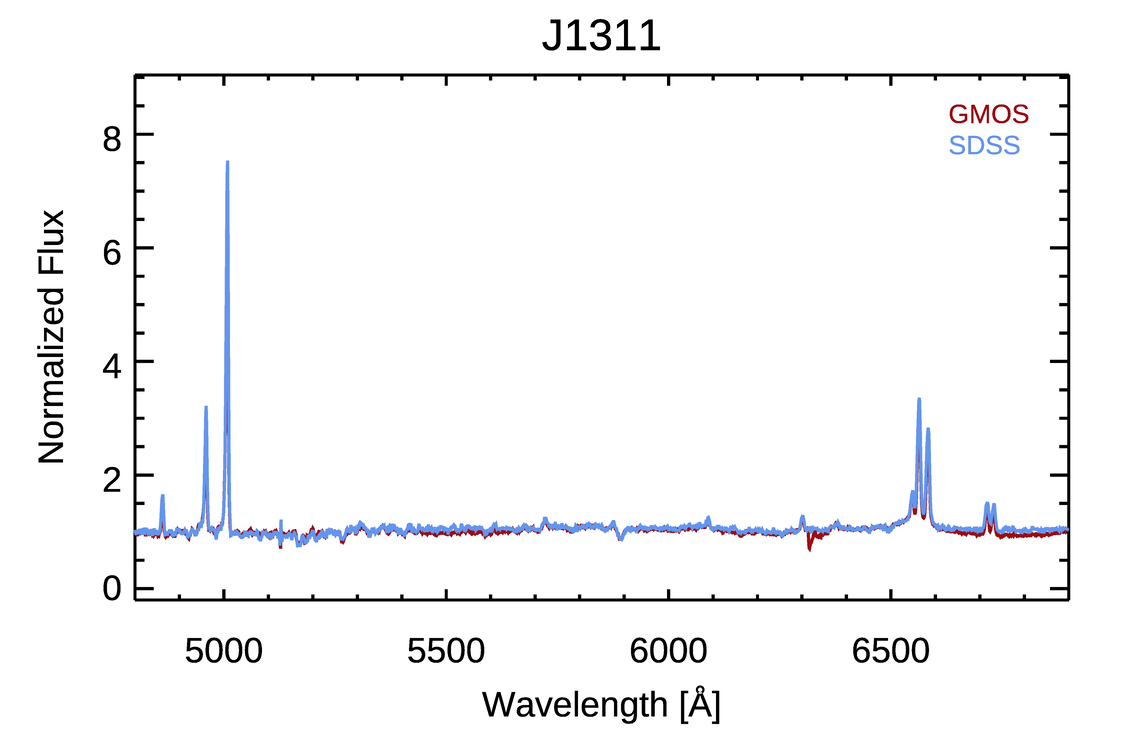}
		\includegraphics[width=0.45\textwidth,angle=0]{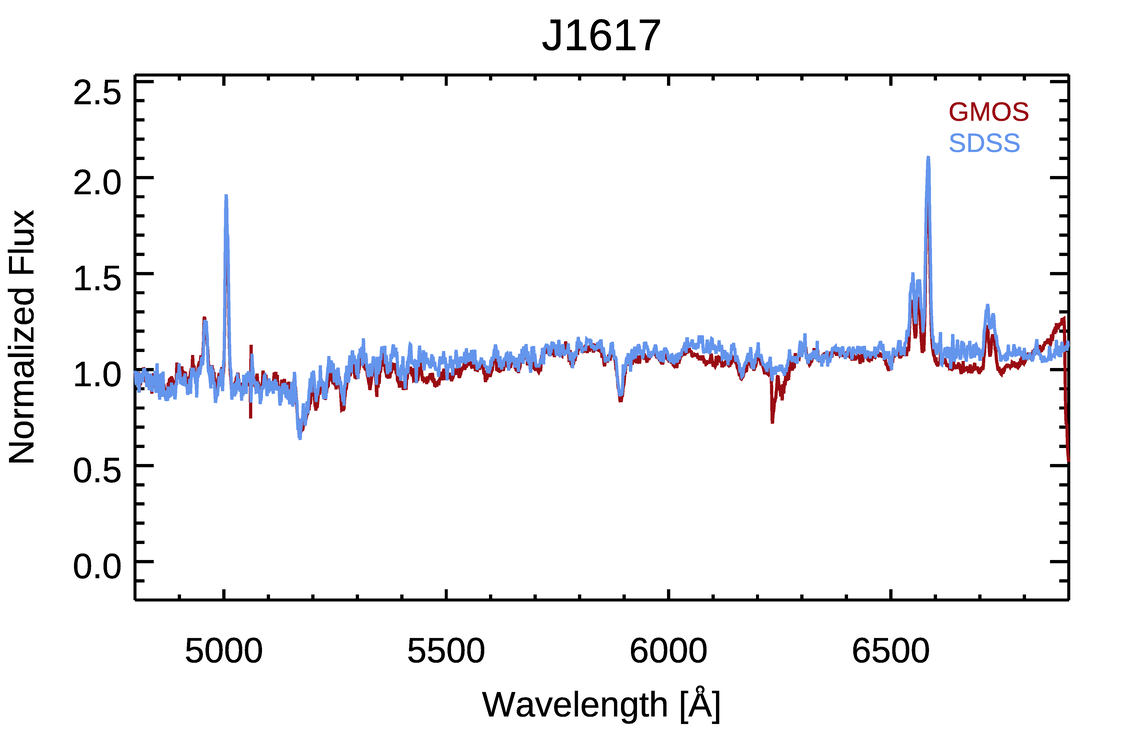}
		\caption{Comparison of the SDSS and GMOS spectra of the sample. GMOS spectra are extracted from the 
			central 3\arcsec spaxels. All the spectra are normalized by the median flux of their continuum.}
		\label{fig:integrated_spec}
	\end{center}
\end{figure*}

\subsection{Observations}
\label{sec:observations}

We used the GMOS-N IFU with the 1-slit mode to observe the sample in 2016A semester (GN-2016A-Q-19, PI: Woo). 
The field of view ($5\farcs0\times3\farcs5$) of GMOS-IFU covered 3-10 kpc scale at the redshifts of our targets, with 
a spaxel scale of $\sim 0\farcs07$ ($\sim$ 60-200 pc). We used the B600 grating and a 2-pixel spectral binning, which 
provides a spectral resolution of R $\sim 1400$ (corresponding to a full width at half-maximum (FWHM) velocity resolution 
of $\sim 215$ km s$^{-1}$) over the spectral range $\sim$ 4500-6800\AA. 
The exposure time of each target ranges from 40.5 to 144 minutes, which were determined based on the SDSS photometry. 
As a part of the K-GMT Science Program, our observations were performed in the priority visitor mode. All targets were observed 
under stable weather conditions with low wind and moderate humidity. For different targets, seeing values varied between 
0\farcs64 and 1\farcs00, corresponding to sub-kpc or kpc spatial resolution (see observation details in Table \ref{tab:sample}).

\section{Data reduction and analysis}
\label{sec:method}

\subsection{Data reduction}
\label{sec:reduction}
We followed the procedure as we previously adopted (\citealt{Karouzos2016,Karouzos2016a,Kang2018}). 
Here, we briefly describe the data reduction processes. We used the Gemini \textit{IRAF} package
to perform the standard procedures for the one-slit mode IFU\footnote{http://www.gemini.edu/sciops/data/IRAFdoc/gmosinfoifu.html}.
We first subtracted the CCD bias from all data frames by using the standard 
bias frame. Then we used the PyCosmic routine (\citealt{Husemann2012}) to remove cosmic rays. Based on the flat-field images 
from the afternoon calibration, we obtained the extraction solution of data frames. The fiber-to-fiber variation is corrected by using 
the twilight flat-field images. Next, we used this solution to extract spectra from the science and calibration frames. We used the 
arc spectra to determine the wavelength solution, which were then used to calibrate the science spectra. The sky background was 
corrected by using the mean sky spectra from the dedicated sky fibers. Finally, we performed flux calibration by using the 
spectrophotometric standard stars and built 3-D data cubes by re-sampling the spaxel size to 0\farcs1$\times$0\farcs1 
(effectively oversampling the data by a factor of $\sim$ 6-10). In Fig. \ref{fig:integrated_spec}, we present the GMOS and 
SDSS spectra for each target. The GMOS spectra are extracted from the central 3\arcsec spaxels for this comparison. 

\begin{figure*}[tbp]
	\begin{center}
		\includegraphics[width=0.45\textwidth,angle=0]{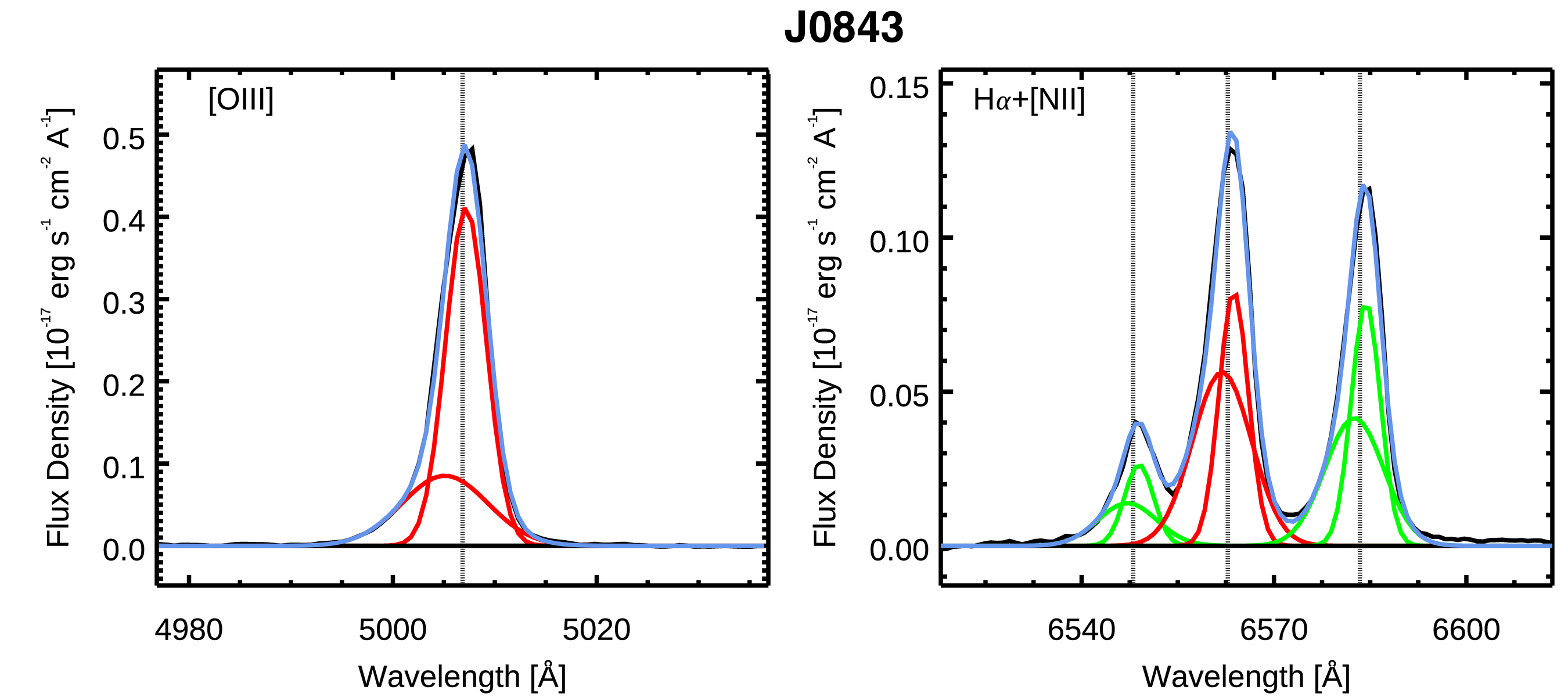}
		\includegraphics[width=0.45\textwidth,angle=0]{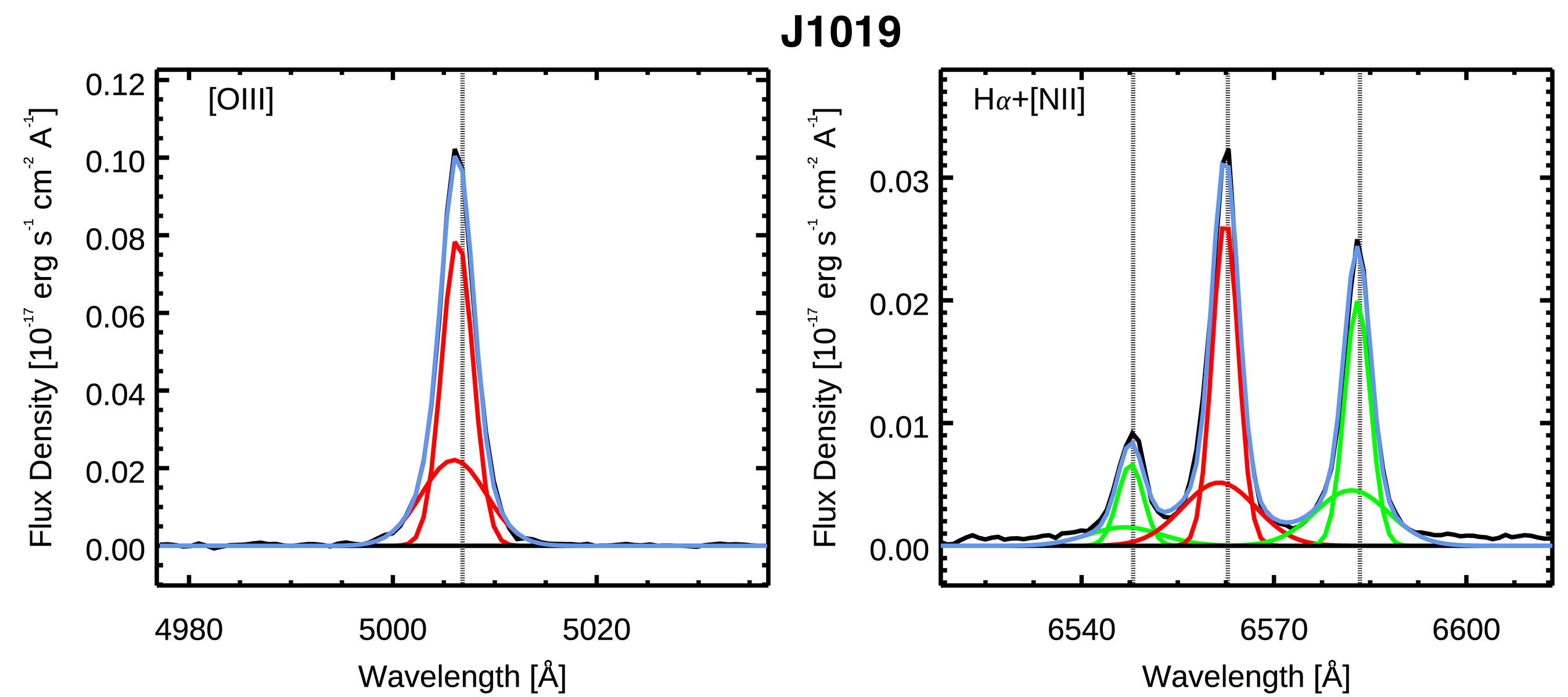}\\
		\includegraphics[width=0.45\textwidth,angle=0]{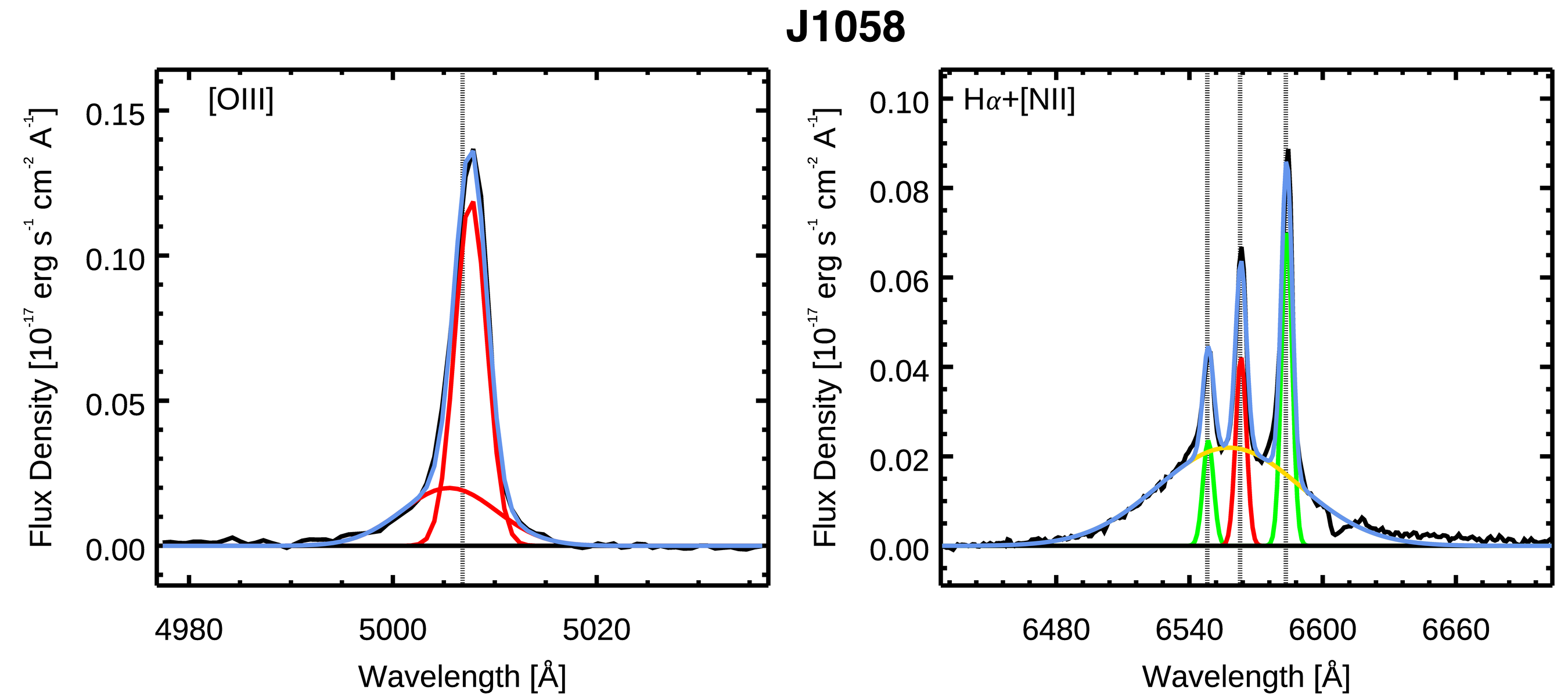}
		\includegraphics[width=0.45\textwidth,angle=0]{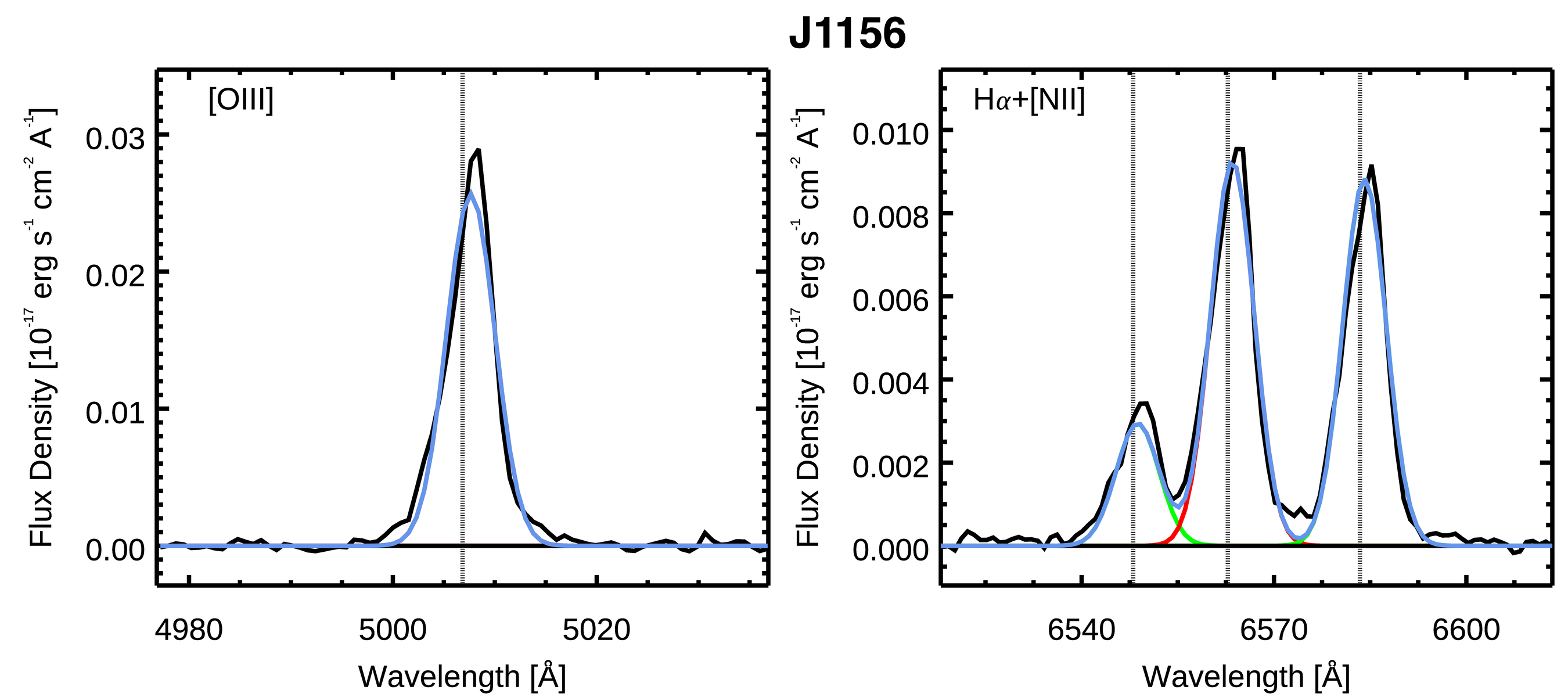}\\
		\includegraphics[width=0.45\textwidth,angle=0]{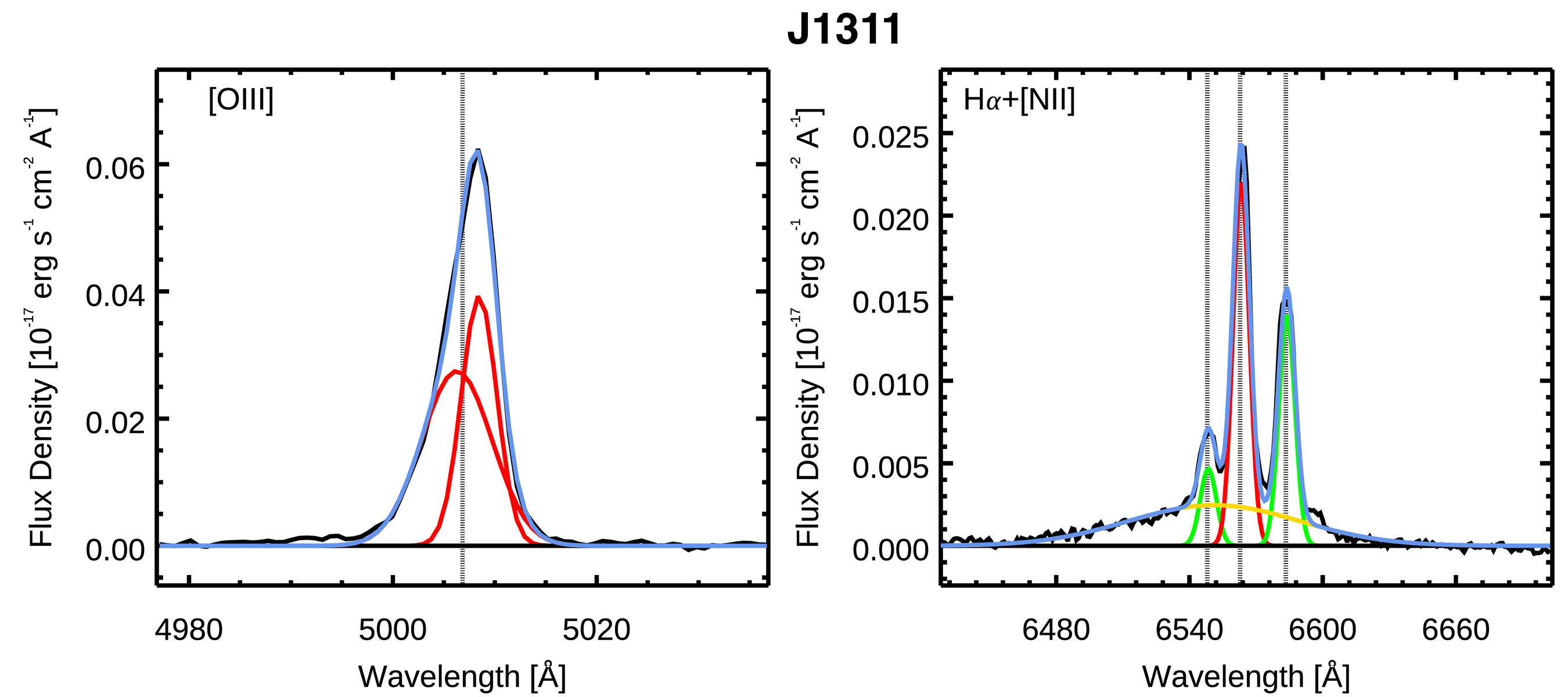}
		\includegraphics[width=0.45\textwidth,angle=0]{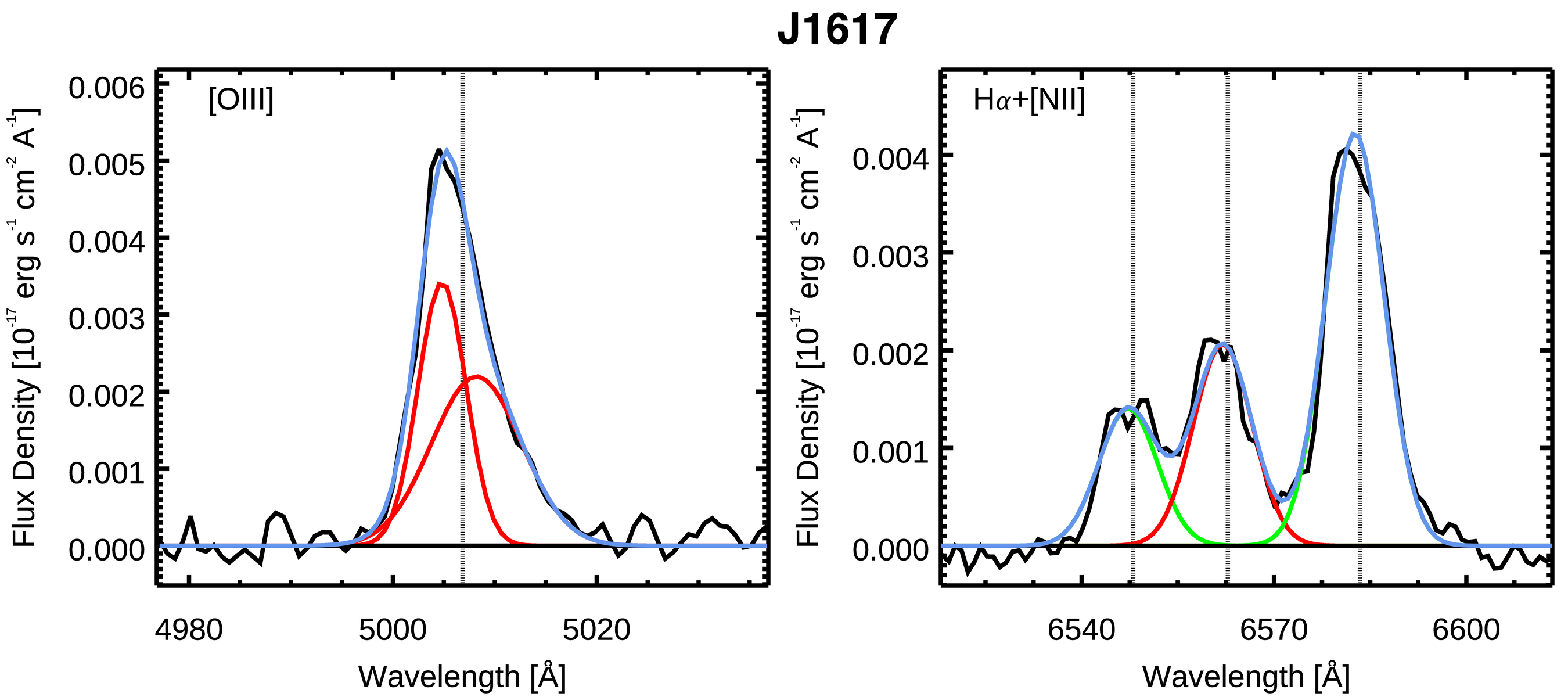}
		\caption{Examples of emission-line fitting of \mbox{[O\,\textsc{iii}]} and H$\alpha$ regions in the central spaxel 
			of each galaxy. The best fit and individual Gaussian components are shown in different colors: blue for the total 
			profile, red for \mbox{[O\,\textsc{iii}]} and H$\alpha$, green for \mbox{[N\,\textsc{ii}]}, and yellow for the very 
			broad H$\alpha$ in J1058 and J1311. The vertical dotted lines indicate the expected center of each line 
			based on the systemic velocity.}
		\label{fig:example_fit}
	\end{center}
\end{figure*}

\subsection{Data analysis}
\label{sec:analysis}
In our previous statistical and IFU-based studies of AGN-driven outflows (e.g., \citealt{Bae2014,Woo2016,Karouzos2016,Karouzos2016a,Woo2017,Bae2017,Kang2017,Kang2018}), 
we have established a set of robust methods to analyze AGN spectra and extract the kinematic properties of ionized gas. With the aim to compare the 
properties of ionized gas in the AGNs with and without strong outflows, we used the same analysis for our sample. 
Here we briefly introduce the methods as below:

First, we used the pPXF \citep{Cappellari2004} to fit the stellar continuum and measure the stellar velocity 
shift (with respect to the systemic velocity) and stellar velocity dispersion for each spaxel. We modeled the continuum with 47 MILES 
simple-stellar population templates, which have solar metallicity, but different ages ranging from 0.63 to 12.6 Gyr \citep{Falcon-Barroso2011}. 
In the fitting process, the systemic velocity of each galaxy was determined from the stellar absorption lines in the spatially integrated 
spectra within the central 3\arcsec\ spaxels.

Second, for each spaxel, we subtracted the best-fitted stellar continuum from the observed spectra to produce the pure emission-line 
spectrum. Then we fitted the H$\beta$, \mbox{[O\,\textsc{iii}]}, \mbox{[N\,\textsc{ii}]}, H$\alpha$, and \mbox{[S\,\textsc{ii}]} 
emission lines, using the Levenberg-Marquardt least-squares algorithm (\citealt{Marquardt1963,More1978}) as implemented 
in the IDL procedure \textit{MPFIT} \citep{Markwardt2009}. We fitted each emission line with up to two Gaussian components and adopted 
an iterative method to determine the number of Gaussian components: (1) The peak amplitude of the broad component should be at least 
larger than 3 times the noise measured at the continuum near the respective emission line. (2) To avoid the detachment of two Gaussian 
components, the distance between their peaks should be smaller than the sum of their widths ($\sigma$). During the fitting of 
H$\alpha$+\mbox{[N\,\textsc{ii}]} and \mbox{[S\,\textsc{ii}]} regions, to reduce the degrees of freedom, we assume the same 
velocity shift (with respect to the systemic velocity) and velocity dispersion for the doublets (\mbox{[N\,\textsc{ii}]} and \mbox{[S\,\textsc{ii}]}), 
while the velocity dispersion is in turn tied to the velocity dispersion of individual H$\alpha$ components. In Fig. \ref{fig:example_fit} we present 
the results of emission-line fitting for the central spaxel of each galaxy.

Third, based on the best-fitted total emission line profile (i.e., either single Gaussian model or the sum of the two Gaussian components),
we calculated the first moment $\lambda_{0}$ and second moment $\Delta\lambda$ for each emission line in each spaxel, which are defined as
\begin{equation}
\label{eq:mom}
\mathrm{\lambda_{0}}=\frac{\int\lambda f_{\lambda}d\lambda}{\int f_{\lambda}d\lambda},\;  \mathrm{\Delta\lambda^{2}}=\frac{\int\lambda^{2} f_{\lambda}d\lambda}{\int f_{\lambda}d\lambda}-\mathrm{\lambda_{0}}^{2}.
\end{equation}
Then we calculated the line flux, the velocity shift (with respect to the systemic velocity) and the intrinsic velocity dispersion.
The instrumental spectral resolution ($\sigma_{inst}\sim91$ km s$^{-1}$) 
is corrected by subtracting it in quadrature from the observed velocity dispersion. 
For some spaxels, emission lines were not resolved and the velocity dispersion of these lines became zero
after correcting for the instrumental spectral resolution. Thus, we set a value of zero for the velocity dispersion in these cases. 
In addition to the best-fitted total line profile, we also performed the above calculations on the profiles of each narrow and 
broad components for two objects, J1019 and J 1058, in which the broad component is relatively well separated from the narrow 
component (see Section 4.3.2). To estimate the uncertainties of the measurements, we performed Monte Carlo simulations and 
produced 100 mock spectra by randomizing the flux using the flux error at each wavelength. We fitted each of these spectra 
and adopted the standard deviation of the measurement distribution as the uncertainty. An iterative 4$\sigma$ clipping algorithm 
was used to exclude the bad fits in this process.

For J1058 and J1311, we find a very broad H$\alpha$ component ($\sigma\sim1400-1700$ km s$^{-1}$) in their spectra, 
which could originate from the broad-line regions (BLRs). Thus, in addition to our regular double Gaussian fitting process, we fit the 
H$\alpha$+\mbox{[N\,\textsc{ii}]} region with one additional broad Gaussian component. The central wavelength and dispersion of this 
component was left to be free during the fitting process. The detailed description of BLRs in J1058 and J1311 will be presented in 
section \ref{sec:BLR_properties}. In our analysis of the other targets, we do not include this broad component because it is assumed 
to present the gas kinematics in the BLRs. 

Based on the above analysis, we obtain the two dimensional maps of continuum flux and emission-line flux, stellar and ionized gas 
velocity and velocity dispersion of each target. To exclude spaxels with weak lines or bad measurements, we employ a S/N limit 
of 3 (based on the peak S/N) for \mbox{[O\,\textsc{iii}]} and H$\alpha$ emission lines. Spaxels with lower S/N are masked as gray 
regions in the two dimensional maps. In section \ref{sec:results}, we will present these results.

\section{Results}
\label{sec:results}

\begin{figure*}[bpt]
\begin{center}
\raisebox{-0.5\height}{\includegraphics[width=0.98\textwidth,angle=0]{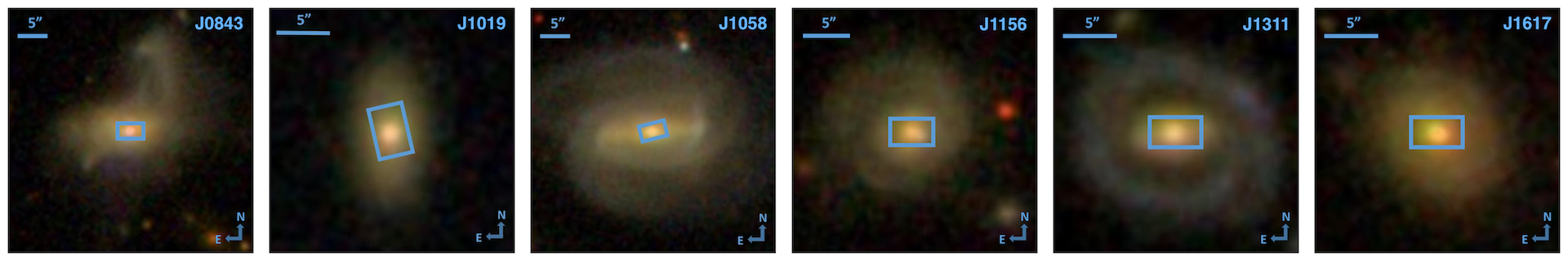}}
\includegraphics[width=1.0\textwidth,angle=0]{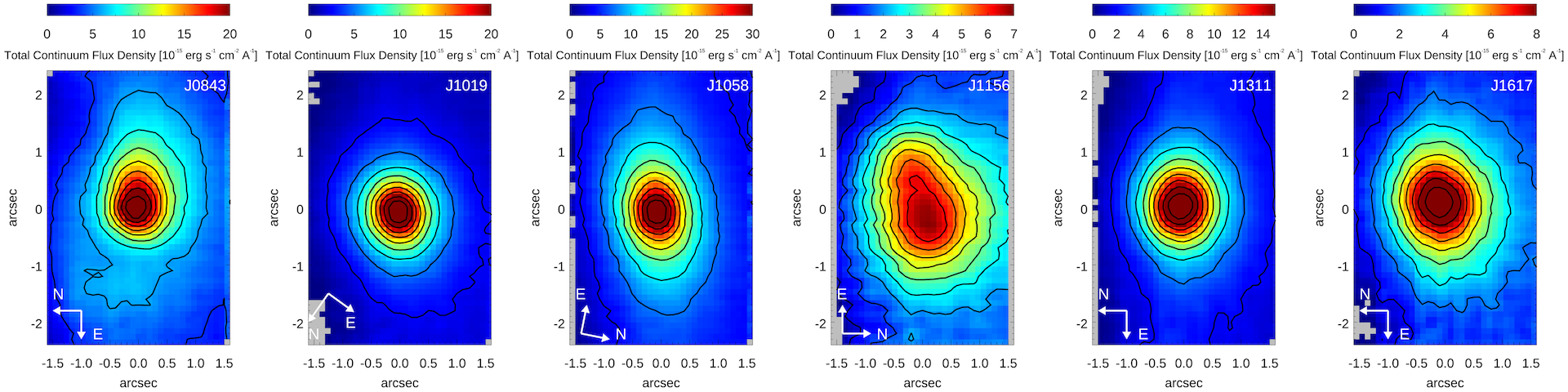}
\includegraphics[width=1.0\textwidth,angle=0]{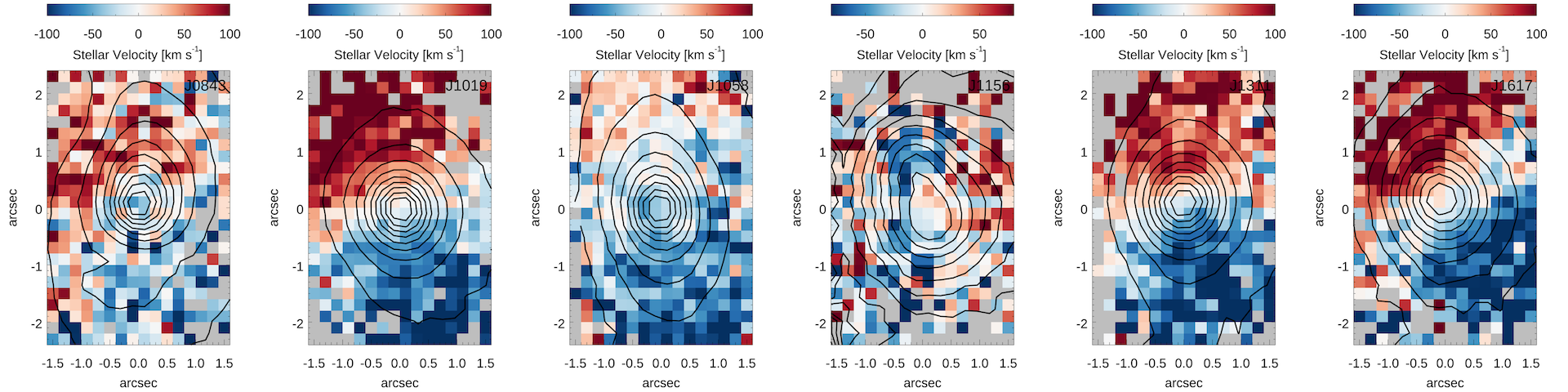}
\includegraphics[width=1.0\textwidth,angle=0]{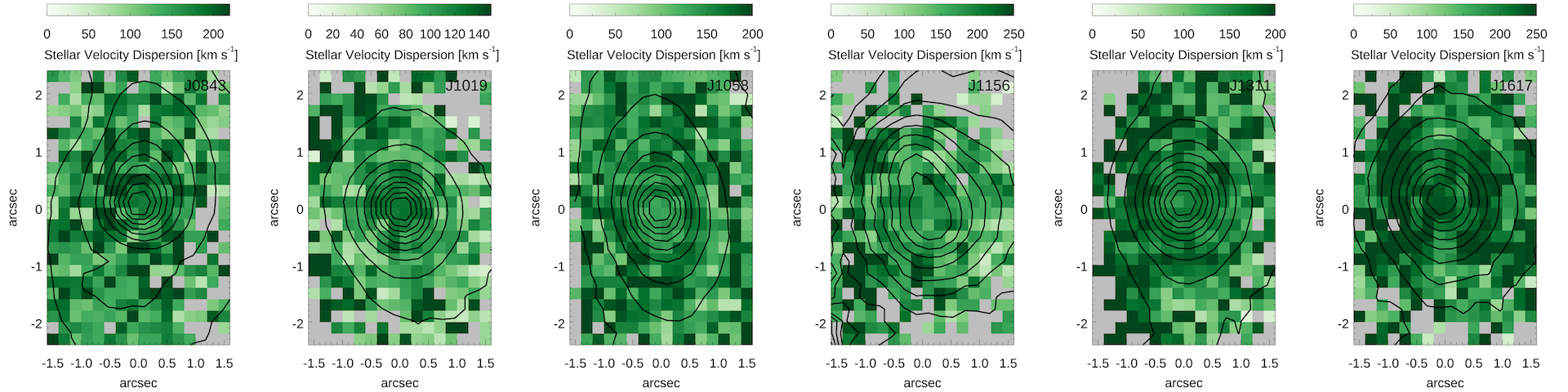}\\
\caption{Top: SDSS \textit{gri} composite images. The FOV of GMOS ($5\farcs0\times3\farcs5$) is shown with a blue box, while the horizontal 
blue bar indicates the 5\farcs0 scale. Second row: Continuum flux maps (integrated over the emission line-free region of the spectrum). Third 
row: Stellar velocity maps. Fourth row: Stellar velocity dispersion maps. The stellar velocity and velocity dispersion are 
obtained from the data cubes with 2 spaxels $\times$ 2 spaxels binning to increase the S/N. Gray region indicates spaxels without reliable 
measurements. From the second to fourth row, the black contours show the continuum flux with 10\% intervals from the peak.}
\label{fig:host_maps}
\end{center}
\end{figure*}

\subsection{Host galaxy properties}
\label{sec:host_galaxy}
In this section, we describe the properties of host galaxies of our sample. More detailed description of individual targets can be found in 
Appendix. In Fig. \ref{fig:host_maps}, we present the SDSS composite images, the continuum flux maps, as well as the stellar velocity 
and velocity dispersion maps. As shown in the SDSS composite images, four targets present almost 
face-on morphology, while J1019 appears moderately inclined (i.e., minor to major axis ratio b/a of 0.60). For J0843, the inclination is uncertain 
due to its disturbed morphology and tidal features, which could be caused by merging process \citep{Lintott2011}.
We detect spiral arms, bars or ring structures in J1058, J1156 and J1311, which are classified as spiral galaxies in the Galaxy Zoo project \citep{Lintott2011}. 
We find no clear structure in J1019 and J1617. J1019 is marked as uncertain morphology in the Galaxy Zoo project, while there is no morphology classification 
of J1617 in the literature. The GMOS continuum flux map of each galaxy is generally consistent with the SDSS images. 

\begin{figure*}[htbp]
	\begin{center}
		\includegraphics[width=1.0\textwidth,angle=0]{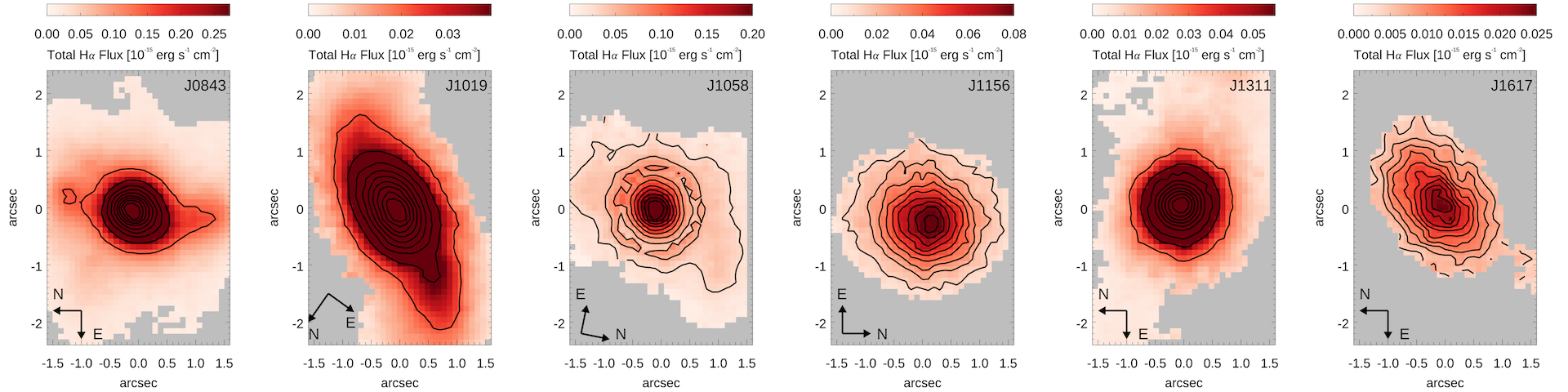}
		\includegraphics[width=1.0\textwidth,angle=0]{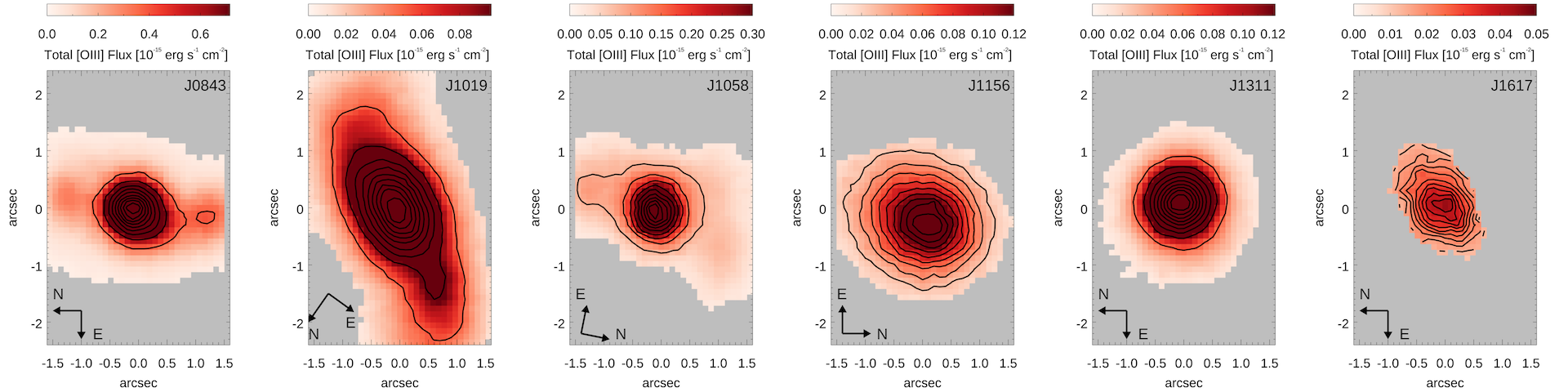}\\
		\caption{H$\alpha$ (top) and \OIII\ (bottom) flux maps. Black contours show the emission line 
			flux with10\% intervals from the peak. Gray spaxels indicate weak or non-detection of the emission lines (i.e., S/N $<$ 3).}
		\label{fig:gas_flux}
	\end{center}
\end{figure*}

For measuring stellar velocity and velocity dispersion, we used 2$\times$2 spaxel binning, in order to increase the S/N 
of the weak stellar lines. The spaxels with failed continuum fitting (e.g. low S/N and unreliable fitting result) are masked as gray regions. 
The instrumental resolution has been corrected for stellar velocity dispersion. In four targets (J1019, J1058, J1311 and J1617), the velocity 
maps show a butterfly shape, suggesting a clear rotation pattern, which is well aligned with the continuum flux distribution. 
J0843 presents a somewhat complex velocity structure, while it shows a velocity gradient along the 
NW-SE direction, suggesting a rotation. In J1156, the velocity map shows no clear pattern. The maximum stellar 
velocity ranges between 180 and 220 km s$^{-1}$ in different targets. The velocity dispersion maps show enhancements in the central parts of 
three galaxies (J0843, J1311, J1617), while there are no significant features in others. The typical stellar velocity dispersion ranges between 100 and 
220 km s$^{-1}$. 

\subsection{Ionized gas flux distribution}
\label{sec:ionized_gas_flux}
\subsubsection{H$\alpha$ and \mbox{[O\,\textsc{iii}]} emission}
In Fig. \ref{fig:gas_flux}, we present the maps of H$\alpha$ and \mbox{[O\,\textsc{iii}]} emission-line flux of our sample. H$\alpha$ and 
\mbox{[O\,\textsc{iii}]} emission lines (with S/N $>$ 3) are detected out to several kpc scales. Considering the seeing (i.e., 0\farcs64 and 1\farcs00, see Table 1), 
the H$\alpha$ and \mbox{[O\,\textsc{iii}]} emission regions are clearly resolved in all targets. Even for J1311, the spatial FWHM of 
\mbox{[O\,\textsc{iii}]} emission is $\sim 0.75$\arcsec, which is still larger than the corresponding seeing size ($0.64$\arcsec) during the 
observation. In three galaxies (J0843, J1058, J1617), the H$\alpha$ emission is slightly more extended than the \mbox{[O\,\textsc{iii}]} 
emission, while in J1019 and J1156, the spatial coverage of H$\alpha$ and \mbox{[O\,\textsc{iii}]} emission are similar. In J1311, the 
H$\alpha$ emission extends to the edge of the GMOS FOV, while the \mbox{[O\,\textsc{iii}]} emission is more concentrated in the 
inner region. Although the spatial coverages of H$\alpha$ and \mbox{[O\,\textsc{iii}]} emission regions varies among the targets,
we find no significant difference in the flux distribution. 

Comparing with the stellar continuum map, the H$\alpha$ and \mbox{[O\,\textsc{iii}]} emission is more concentrated 
in the inner region. In four targets (J1019, J1156, J1311, J1617), the spatial distribution of H$\alpha$ and \mbox{[O\,\textsc{iii}]} 
emission generally follows that of the stellar continuum. In contrast, J0843 and J1058 show large difference of the flux distribution between
ionized gas and stars, presumably due to the complex gas kinematics in these galaxies
(see Section \ref{sec:ionized_gas_kinematics} for details).

\begin{figure}[htbp]
\begin{center}
\includegraphics[width=0.45\textwidth,angle=0]{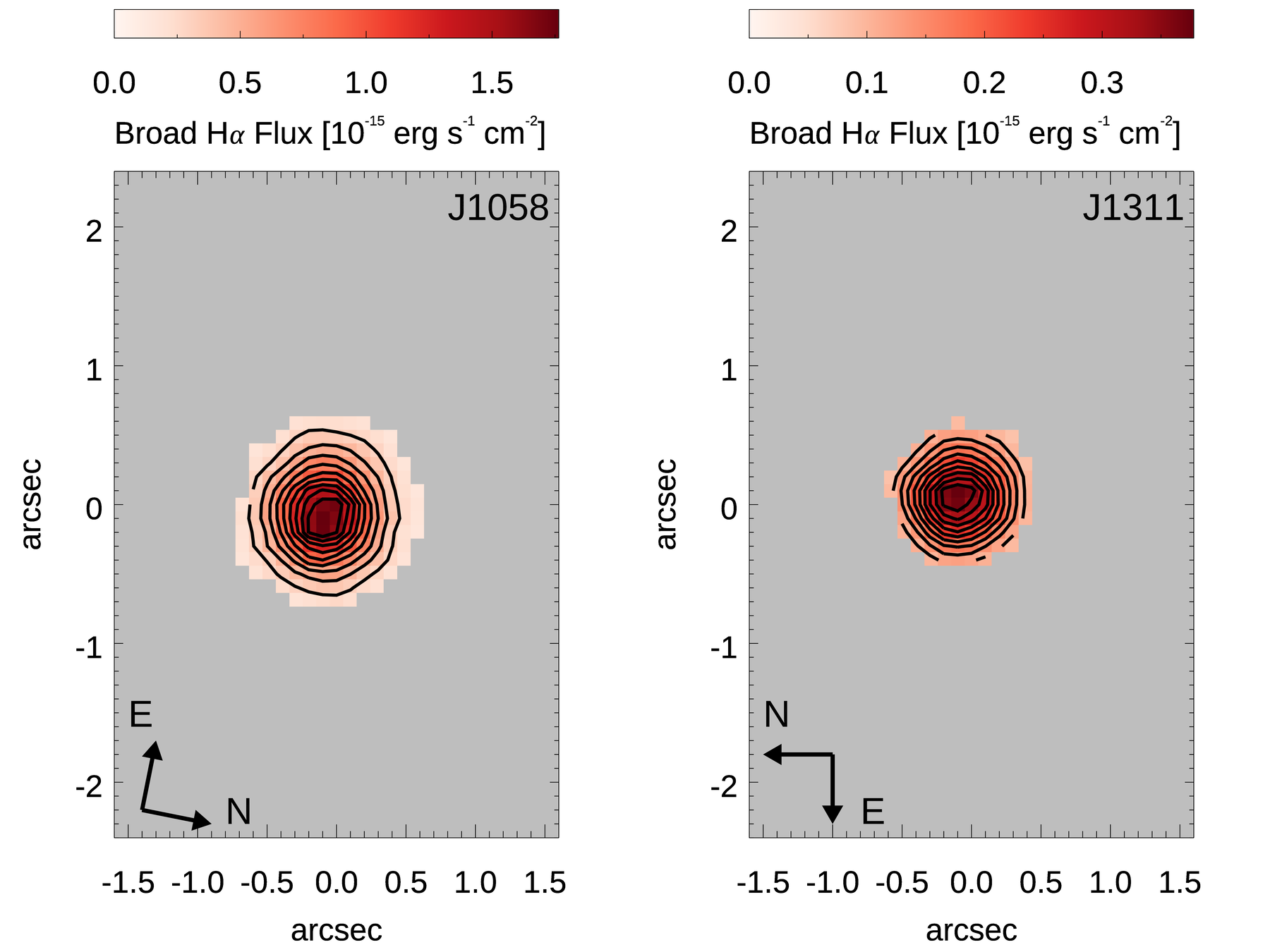}
\caption{Flux maps of the very broad H$\alpha$ component in J1058 and J1311. Black contours show the emission line flux with 10\% intervals from 
the peak. }
\label{fig:broad_ha_flux}
\end{center}
\end{figure}

\subsubsection{H$\alpha$ emission from the broad line region}
\label{sec:BLR_properties}
As described in section \ref{sec:analysis}, we find a very broad H$\alpha$ component in the spectra of J1058 and J1311, which is 
originated from the BLR. This broad component in H$\alpha$ is present within the central $\sim 1.5$\arcsec\ scale (see Fig. \ref{fig:broad_ha_flux}). 
The size of the very broad H$\alpha$ emission is FWHM $\sim 0.64$\arcsec, representing the seeing size during the observation. The velocity and 
velocity dispersion of this component are almost constant within the detected region, confirming that the broad H$\alpha$ emission is a point source 
in our observation with a limited spatial resolution. The velocity shift of this component is $\sim -230$ km s$^{-1}$ and $-540$ km s$^{-1}$, while the 
velocity dispersion is $\sim$ 1400 km s$^{-1}$ and 1700 km s$^{-1}$, respectively, for J1058 and J1311. Using the luminosity of this H$\alpha$ 
component and the scaling relation calibrated by \citet{Woo2015}, we estimate the black hole mass as $2.6 \times 10^6$ \(M_\odot\) and 
$4.1 \times 10^6$ \(M_\odot\), respectively for J1058 and J1311.

\subsection{Ionized gas kinematics}
\label{sec:ionized_gas_kinematics}
\subsubsection{Velocity and velocity dispersion maps}

\begin{figure*}[tbp]
	\centering
		{\Large Velocity}\\
		\begin{flushleft}
		\includegraphics[width=1.0\textwidth,angle=0]{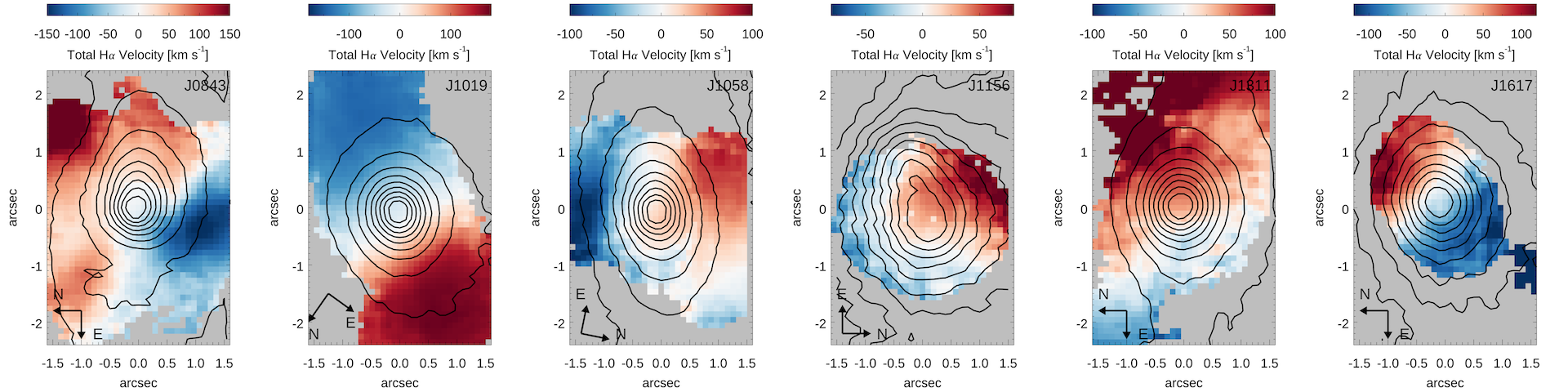}
		\includegraphics[width=1.0\textwidth,angle=0]{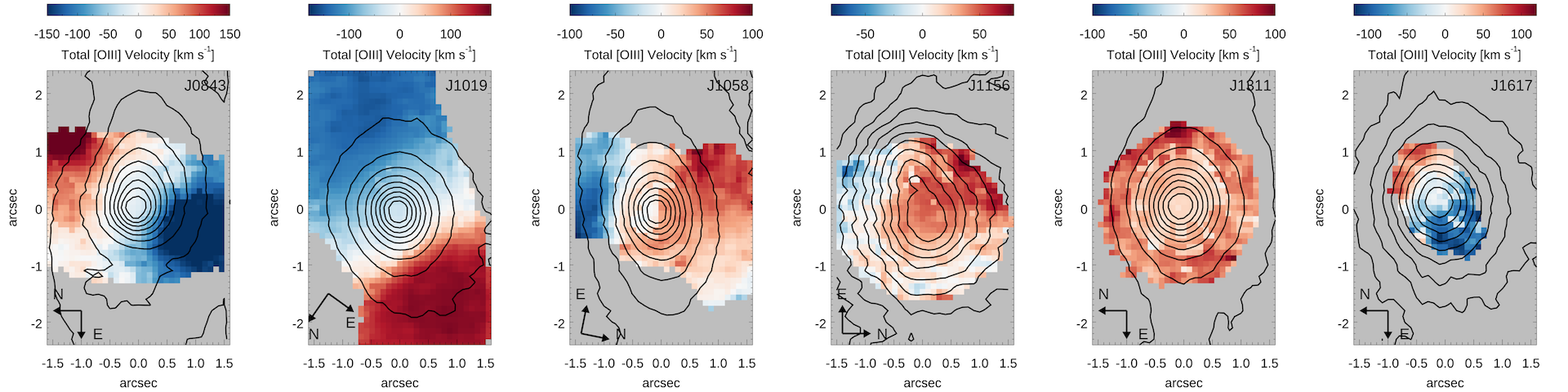}\\
		 \end{flushleft}
		{\Large Velocity dispersion}\\
		\begin{flushleft}
		\includegraphics[width=1.0\textwidth,angle=0]{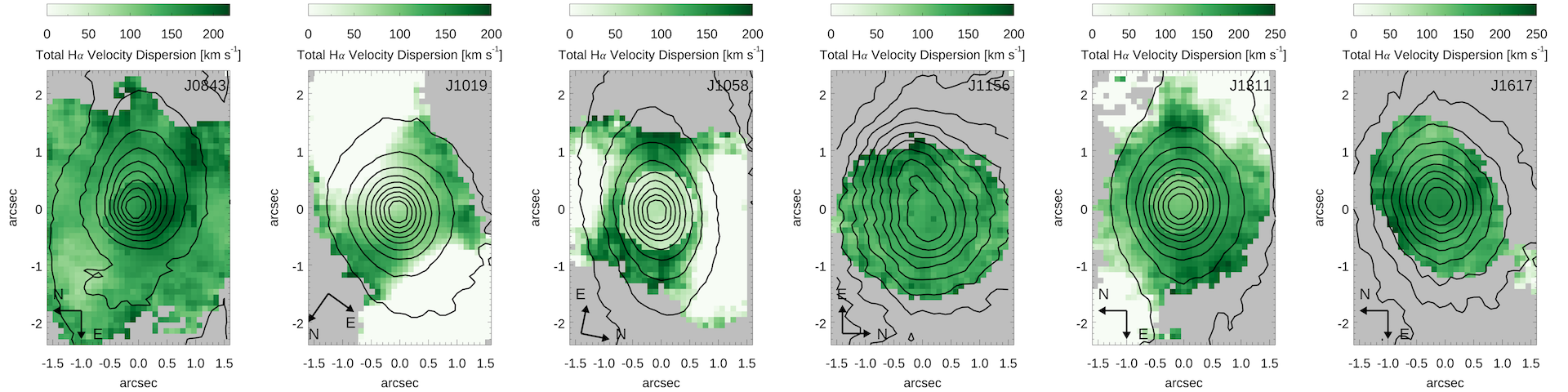}
		\includegraphics[width=1.0\textwidth,angle=0]{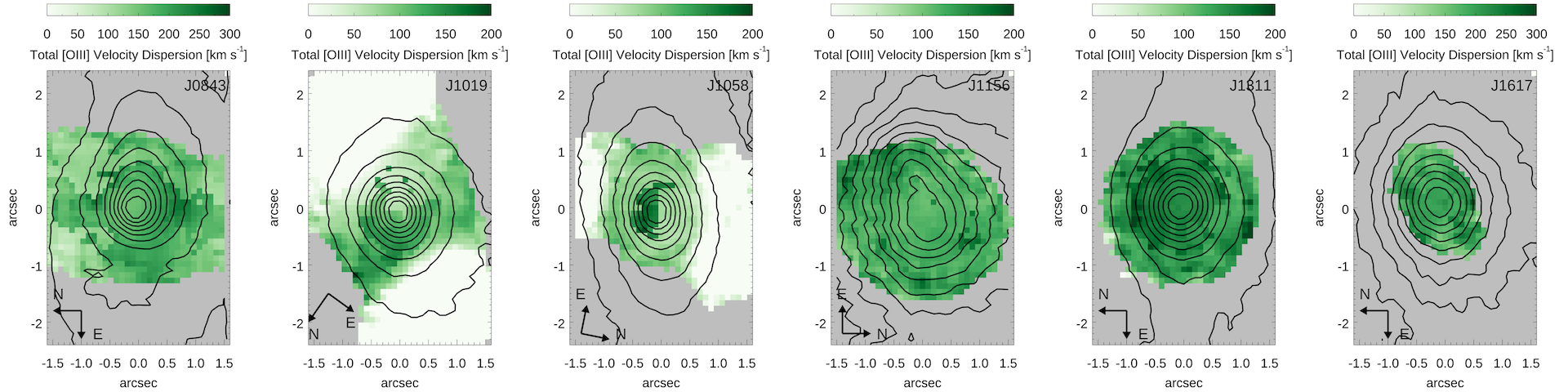}\\
		 \end{flushleft}
		\caption{Velocity and velocity dispersion maps derived from the best-fitted total profile of H$\alpha$ and \mbox{[O\,\textsc{iii}]} emission lines. 
			As in Fig. \ref{fig:host_maps}, black contours indicate the stellar continuum flux with10\% intervals 
			from the peak. Gray spaxels indicate weak or non-detection of the emission lines (i.e., S/N $<$ 3).
			Note that the velocity of ionized gas is measured with respect to the systemic velocity based on stellar absorption lines at the center of each galaxy.}
		\label{fig:gas_kin_t}
\end{figure*}

\begin{figure*}[bpt]
	\begin{center} 
		\includegraphics[width=0.80\textwidth,angle=0]{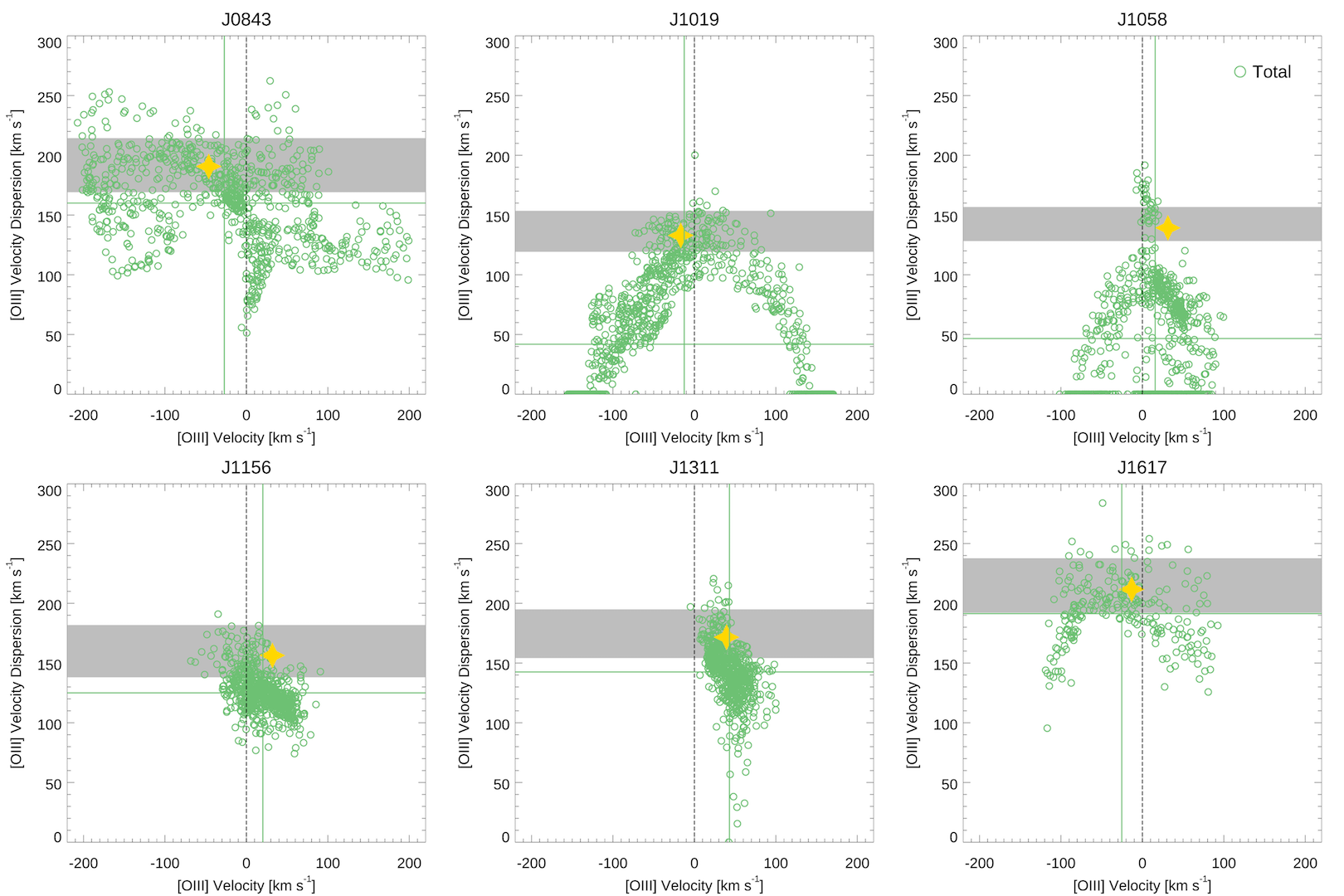}\\
		\caption{VVD diagram based on the best-fitted total profile of \mbox{[O\,\textsc{iii}]} in each spaxel, along with the mean values (green lines). 
				Gray-shaded stripes denote stellar velocity dispersion with a 3$\sigma$ uncertainty while gold stars indicate \mbox{[O\,\textsc{iii}]} velocity and 
				velocity dispersion measured from the SDSS spectra.}
		\label{fig:all_vvd_oiii}
	\end{center}
\end{figure*}

In Fig. \ref{fig:gas_kin_t}, we present the velocity and velocity dispersion maps derived from the best-fitted total line profile 
of H$\alpha$ and \mbox{[O\,\textsc{iii}]}. The spaxels with weak lines or bad measurements (i.e., S/N $<$ 3 for H$\alpha$ 
and \mbox{[O\,\textsc{iii}]} emission lines) are masked as gray regions in these maps. Comparing with stars, the ionized gas 
presents complex kinematic signatures. We describe the detailed properties of each target as below.

\textbf{J0843:} The kinematics of ionized gas, particularly \Ha, seems affected by the merging process (see also the velocity-velocity dispersion diagram in Fig. \ref{fig:all_vvd_oiii}), 
while both H$\alpha$ and \mbox{[O\,\textsc{iii}]} velocity maps show a rotation pattern in the NW-SE direction, which is weakly present in the stellar velocity map. 
While it is possible that AGN-driven outflows influence gas kinematics, but the complex nature of gas motion is probably due to the merging process. 

\textbf{J1019:} The velocity maps of the ionized gas show an opposite spatial trend compared to the stellar velocity map. 
One possible explanation is that the ionized gas is counter-rotating compared to stars.
The counter-rotating gas can be present due to external processes, e.g. major mergers, minor mergers or gas accretion (\citealt{Corsini2014}). 
This phenomenon is more often detected in elliptical and lenticular galaxies, while it is relatively rare in late type galaxies (\citealt{Corsini2014,Chen2016}). 
By examining the large scale environment using the SDSS image, we find no nearby companion galaxy, suggesting that gas supply from a nearby 
companion is unlikely.  The bi-conical outflows driven by AGN is consistent with the observed gas kinematics, if the outflow direction is along the 
kinematic major axis. As predicted by the 3D bi-conical model of AGN-driven outflows (Shin et al. in prep.), the gas velocity dispersion will be enhanced in the 
central part of the bi-cone, due to the combined effect of gas outflows and the point spread function (PSF) smearing effect. The observed velocity dispersion 
of \OIII\ and \Ha\ is consistent with this prediction. As shown in the velocity dispersion maps, the ionized gas has low velocity dispersion along the kinematic 
major axis, while velocity dispersion is significantly enhanced in the central region. Note that due to the limited spectral resolution, emission lines are not 
resolved in the outer pixels (i.e., velocity dispersion is set to 0). As the broad and narrow components of \OIII\ are relatively well separated in the line profile, 
we also present the VVD diagram of these components in Fig. \ref{fig:two_vvd_oiii} (see Section 4.3.2).

\textbf{J1058:} \OIII\ and \Ha\ velocity maps show that the kinematic major axis of the ionized gas is misaligned with respect to 
that of stars, by 40 degree. The misalignment can be caused by internal processes as well as external processes. While a mild 
misalignment can be caused by the effect of non-axisymmetric structures (e.g., bars, spiral arms) and decoupled stellar components, 
a large misalignment is mainly considered as the result of galaxy interactions or gas accretion (\citealt{Dumas2007,Davies2014,Jin2016}). 
From the large scale SDSS image, we find no nearby companion of this galaxy. Considering the clear and strong bar structure in J1058, 
we consider that the gravitational perturbation of large-scale bar may be responsible for producing the kinematic misalignment between gas and stars.
However, as predicted by the hydrodynamical simulation of gas flows within a bar structure \citep{Li2015}, the gas velocity dispersion 
will be enhanced along the leading or trailing side of the bar. The observed velocity dispersion of the ionized gas contradicts to 
this prediction. The significant enhancement of velocity dispersion can only be observed in the central region of the galaxy, while it is 
very low along the kinematic major axis of the ionized gas. The bi-conical outflows driven by AGN can provide 
a better explanation. The outflow direction is independent of the axis of the stellar disk, and the velocity map may represent the bi-conical outflows
in the N-S direction. The enhancement of velocity dispersion at the center of the bi-cone can be also naturally explained by the overlap of the 
approaching and receding cones due to the PSF smearing effect. Note that H$\alpha$ velocity dispersion is relatively low at the central region. 
However, this may be an artifact as the line profile of H$\alpha$ in the central spaxels is not very well constrained because of the presence of 
the very broad H$\alpha$ component (see Fig. \ref{fig:broad_ha_flux}).

\textbf{J1156:} \OIII\ and \Ha\ velocity maps show an irregular pattern with positive velocity (i.e., redshift) in the central region, while it could be 
interpreted as a weak rotation. Note that stars do not show a clear rotation pattern. The velocity is limited within $\pm$100 \kms, while the typical 
velocity dispersion is $\sim$ 120 km s$^{-1}$. These kinematic properties can be interpreted as due to the face-on orientation of a rotating disk of 
a relatively low mass galaxy although we cannot rule out that the kinematic pattern is due to outflows.

\textbf{J1311:} There is clear difference between velocity maps of \OIII\ and \Ha. While H$\alpha$ follows the 
stellar rotation pattern, \mbox{[O\,\textsc{iii}]} shows relatively weak positive velocities (i.e., redshifts) without a clear rotation pattern. The bi-conical 
outflows can explain the positive velocities of \OIII, if the angle between the bi-cone axis and the dusty galactic plane is small and the approaching 
cone is obscured. However, \OIII\ velocity is relatively small ($<$ 100 \kms) and \OIII\ velocity dispersion is also comparable to stellar velocity dispersion. 
Thus, we have no strong evidence of AGN-driven outflows. The smooth distribution of \OIII\ velocity dispersion may reflect the PSF smearing effect 
of the centrally-concentrated small scale NLR. 

\textbf{J1617:} Both H$\alpha$ and \mbox{[O\,\textsc{iii}]} velocity maps show a rotation pattern, which is consistent with the stellar velocity map. 
The typical velocity and velocity dispersion of the ionized gas are also comparable with those of stars. We find no clear evidence of outflows.

In summary, the kinematics of the ionized gas in our sample are governed by various physical processes, including merging, 
AGN-driven outflow and host galaxy gravitational potential. In two targets, namely, J1019 and J1058, the ionized gas kinematics is significantly 
different from that of stars, which is consistent with the bi-conical outflows driven by AGN. 

\subsubsection{Velocity-velocity dispersion diagram}

In Fig. \ref{fig:all_vvd_oiii}, we present the velocity versus velocity dispersion (VVD) diagram of each target. 
As described by \citet{Karouzos2016}, the outflow components are often blueshifted and broad in the case 
of AGNs with strong outflows. Thus, the spaxels with outflow signatures will be located at the upper left corner of the VVD diagram. 
In contrast, we find no such a trend in our sample, except for J1058, for which the comparison of gaseous and stellar kinematics indicates 
AGN-drive outflows (S.4.3.1). In J0843, the VVD diagram shows no regulated pattern, reflecting the effect of merging on the gas kinematics. 
While two targets, J1156 and J1311 show neither a rotation pattern or strong outflow signatures, J1617 shows a rotation pattern without outflow signatures. 

\begin{figure}[bpt]
	\begin{center} 
		\includegraphics[width=0.35\textwidth,angle=0]{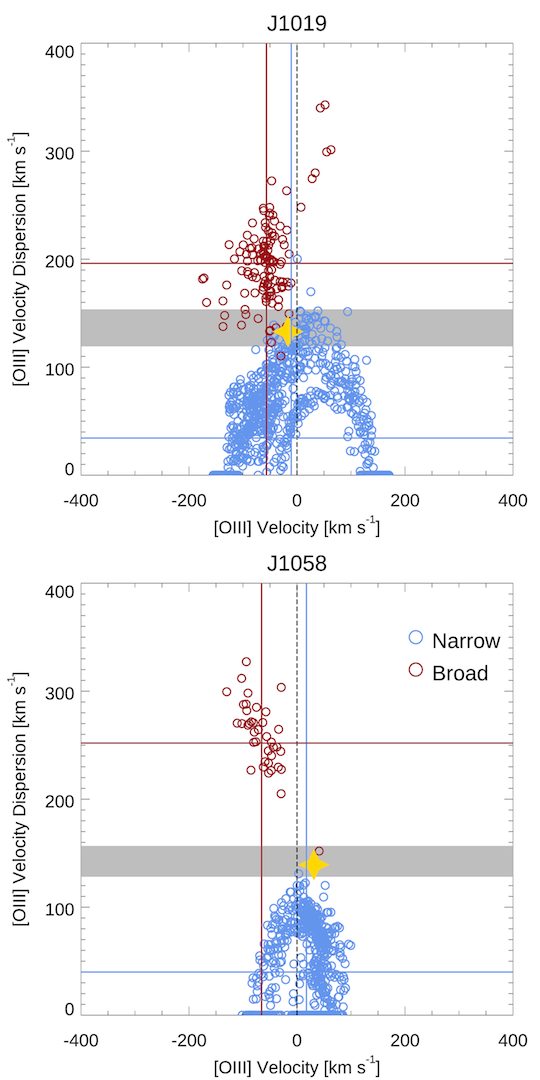}\\
		\caption{VVD diagram of the broad (red) and narrow (blue) components of \OIII\ for J1019 (top) and J1058 (bottom), along with the mean values (red and blue lines). 
			Gray-shaded stripe denotes stellar velocity dispersion with a 3$\sigma$ uncertainty. Gold stars indicate\mbox{[O\,\textsc{iii}]} velocity and velocity dispersion measured from the SDSS spectra.}
		\label{fig:two_vvd_oiii}
	\end{center}
\end{figure}

For J1019 and J1058, we also present the VVD diagram, respectively, using the narrow and broad components of \OIII\ in Fig. \ref{fig:two_vvd_oiii}, 
since these two objects have outflow signatures in the velocity maps. Compared to AGNs with strong outflows, the VVD diagrams of these two 
objects show similar patterns with the broad component extending to the upper left (i.e., high velocity dispersion and high negative velocity), 
which further supports the presence of AGN-driven outflows. However, the broad component of \OIII\ has a mean velocity dispersion larger 
than stellar velocity dispersion by a factor of 1.4-1.8, indicating that \OIII\ velocity dispersion is not as high as those of AGNs with strong 
outflows \citep{Karouzos2016}. These results suggest that relatively weak outflows are present in J1019 and J1058.

\subsection{Photoionization properties}
\label{sec:ionized_gas_excitation}

\begin{figure*}[tbp]
	\begin{center}
		\includegraphics[width=0.75\textwidth,angle=0]{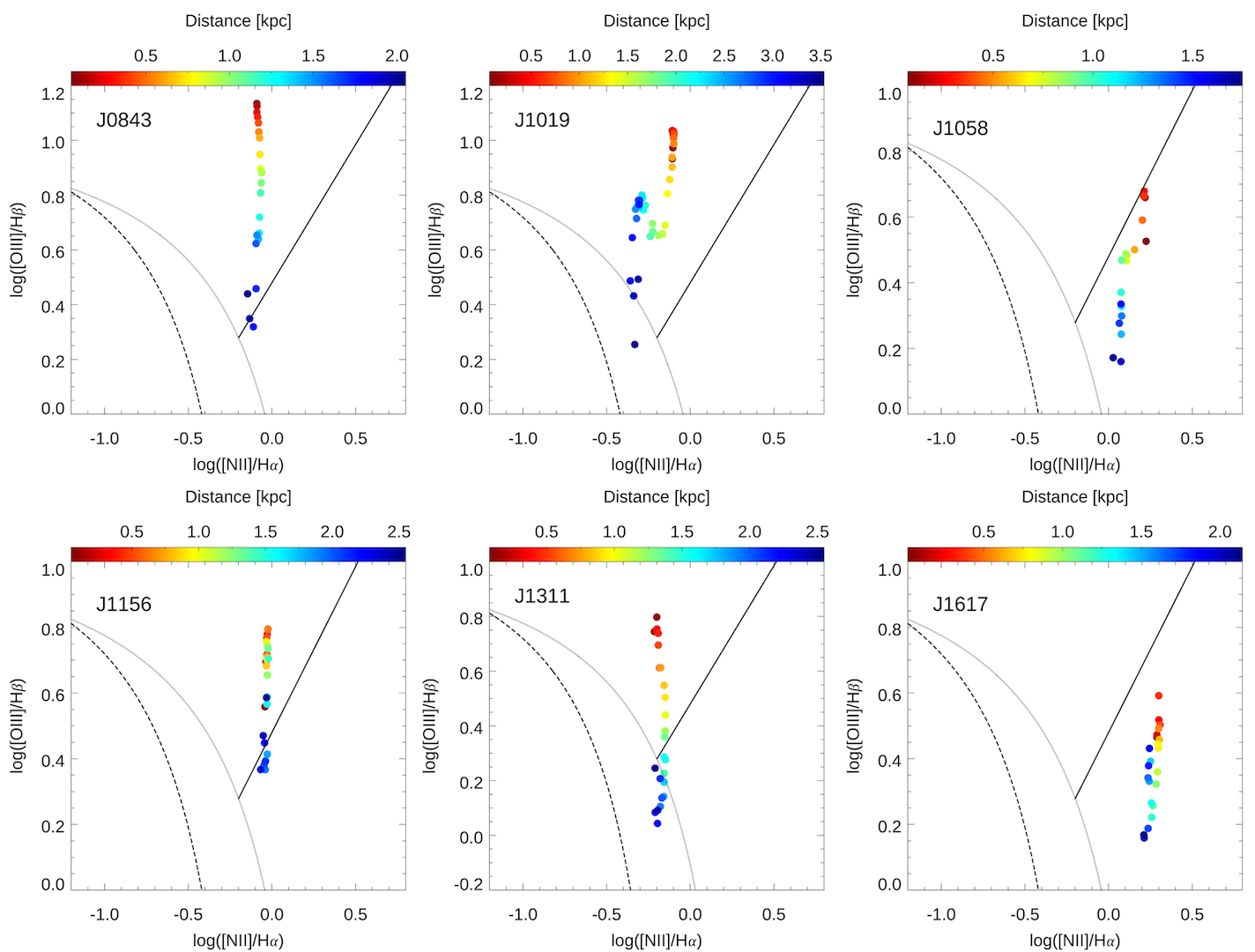}
		\caption{Spatially resolved BPT diagrams.
				The mean flux ratios of \OIII/H$\beta$ and \NII/H$\alpha$ are calculated within the distance bin of 0.1 kpc and color-coded accordingly. 
				The dotted and dashed curves indicate the demarcation lines defined 
				by \citet{Kewley2001} and \citet{Kauffmann2003}, respectively. The solid line shows the Seyfert/LINER demarcation of 
				\citet{CidFernandes2010}. }
		\label{fig:bpt_diagrams}
	\end{center}
\end{figure*}

\begin{figure*}[tbp]
	\begin{center}
		\includegraphics[width=0.5\textwidth,angle=0]{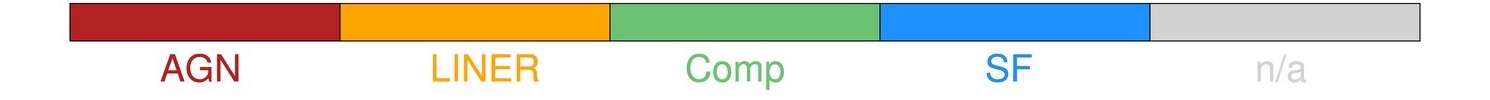}
		\includegraphics[width=0.9\textwidth,angle=0]{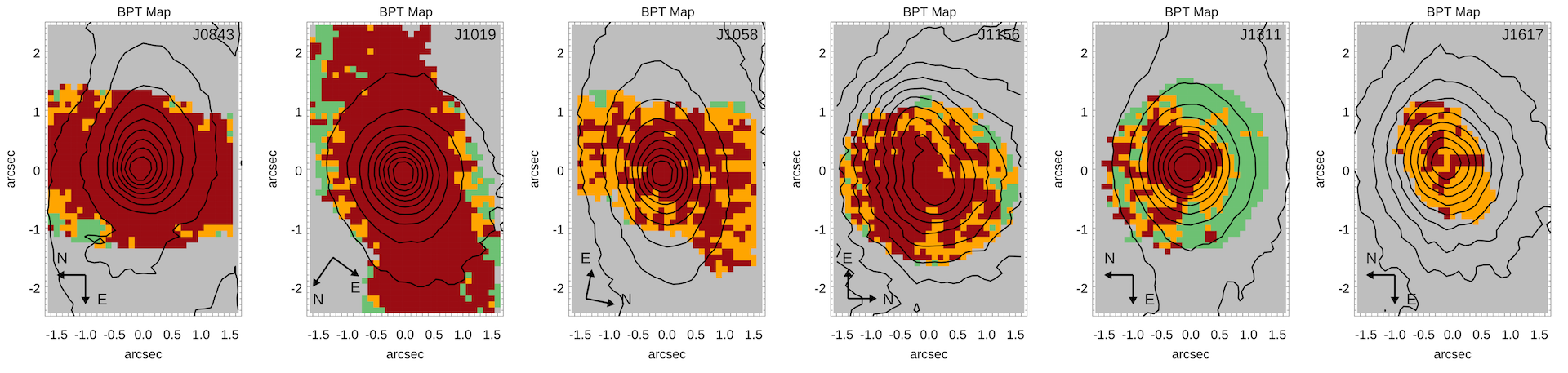}
		\caption{BPT classification maps.
			As in Fig. \ref{fig:host_maps}, black contours indicate the the stellar continuum flux at 10\% intervals from the peak. Gray spaxels 
			indicate spaxels without classification due to weak or non-detection of the emission lines (i.e., S/N $<$ 3).}
		\label{fig:bpt_maps_t}
	\end{center}
\end{figure*}

\begin{figure*}[tbp]
	\begin{center}
		\includegraphics[width=0.9\textwidth,angle=0]{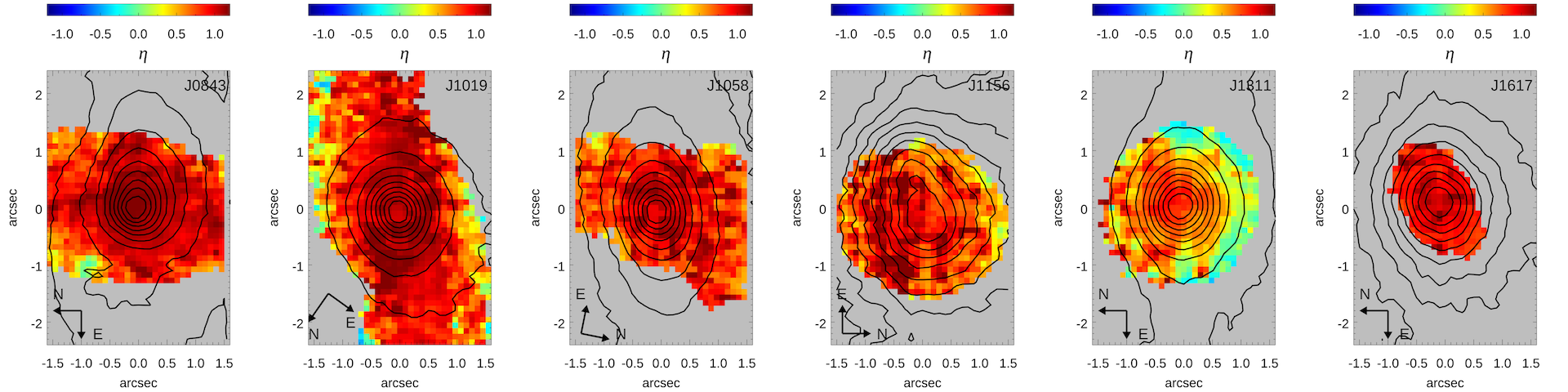}
		\caption{Photoionization classification maps.
			We define the $\eta$ parameter following the scenario of \citet{Erroz-Ferrer2019} and use different colors to indicate its value across the galaxies.
			As in Fig. \ref{fig:host_maps}, black contours indicate the the stellar continuum flux at 10\% intervals from the peak. Gray spaxels 
			indicate spaxels without classification due to weak or non-detection of the emission lines (i.e., S/N $<$ 3). } 
		\label{fig:bpt_maps}
	\end{center}
\end{figure*}

\subsubsection{Spatially resolved BPT diagram}

By combining the flux ratios of \OIII/H$\beta$ and \NII/H$\alpha$, we present the Baldwin, Phillips, \& Terievich (BPT) 
diagram \citep{Baldwin1981,Veilleux1987} to investigate the source of ionization. We adopt the criteria 
from \citet{Kewley2001} and \citet{Kauffmann2003} to classify the AGN, composite, and star-forming regions. 
For dividing Seyfert and LINER regions, we use the demarcation of \citet{CidFernandes2010}.
In Fig. \ref{fig:bpt_diagrams}, we present the BPT classification of each spaxel as a function of distance from the center. 
Note that we measure the flux of each emission line using the best-fitted total line profile. We employ a S/N limit of 3 for 
the \mbox{[O\,\textsc{iii}]}, \NII\ and H$\alpha$ emission lines, whereas we relax 
the S/N limit to 1 for the H$\beta$ line. Thus, BPT classification is not available in the outer pixels if not all four emission lines are detected. 
In order to trace the radial change of the \OIII/H$\beta$ and \NII/H$\alpha$ flux ratios, we calculate the mean flux 
ratios within the distance bin of 0.1 kpc in the projected plane.

In all targets, we observe that the majority of spaxels is classified as Seyfert/LINER region, while there is a clear 
radial trend from AGN-dominated photoionization toward
composite region. This radial change is mainly due to the decrease of the \OIII/H$\beta$ flux ratio.
A significant change of the \NII/H$\alpha$ flux ratio is detected in J1019 and J1058, which are the two AGNs with outflows. 
The origin of this sudden change is unclear and it may be due to the change of gas metallicity or ionization condition.

\subsubsection{BPT morphology}

In Fig. \ref{fig:bpt_maps_t}, we present the photoionization classification maps based on the location of spaxels in the BPT diagram. 
Most targets show AGN-dominated photoionization at the center. In two objects, J1019 and J1311, the AGN-dominated region is surrounded by
composite region, which is consistent with our previous finding of a ring-like structure of star-forming region around the AGN-dominated region
\citep{Karouzos2016, Kang2018}. 

In order to investigate the spatial change of the line ratios, we introduce the $\eta$ parameter following the scheme of \citet{Erroz-Ferrer2019}. 
We use this parameter to describe the variation of the ionization level, quantifying the contribution from AGN photoionization. 
$\eta$ is calculated as the orthogonal distance from the bisector of the two demarcation lines: the line between AGN region and composite 
region \citep{Kewley2001} and the line between star-forming region and composite region \citep{Kauffmann2003}. We normalize $\eta$ to be 
equal to 0.5 at the demarcation line of \citet{Kewley2001} and -0.5 at the demarcation line of \citet{Kauffmann2003}.
In four targets (i.e., J0843, J1019, J1058, J1156), the $\eta$ is greater than 0.5 in most spaxels, in which four emission lines are detected, while 
the central several kpc area shows the largest $\eta$ value, indicating the dominance of AGN photoionization. The lower $\eta$ in the outer part 
of the emission regions indicates the decrease of AGN contribution. At the edge of the AGN-dominant region, the $\eta$ becomes lower than 0.5, 
suggesting a mix of the photoionization from AGN and star-formation. In J1311, AGN photoionization is mainly dominated in the central region, 
while a ring-like structure with $\eta$ less than 0.5 is present in the circumnuclear region. In the case of J1617, the whole emission region is 
dominated by the AGN photoionization. As shown in Fig. \ref{fig:bpt_maps_t}, however, the line flux ratios indicate LINER-like emission except for the central spaxels.

Comparing with the BPT morphology of AGNs with and without strong outflows, we find no significant difference. Based on the BPT morphology 
of 6 AGNs with strong outflows, \citet{Karouzos2016a} concluded that all objects present the signs of circumnuclear star formation \citep[see also][]{Kang2018}. 
AGNs without strong outflows in our sample also show similar morphology with an AGN-dominant center and a mixing zone of AGN and star-formation. 
Due to the lack of enough S/N ratios in the outer part of the FOV, it is difficult to conclude whether the ring-like structure of star-forming region 
is present for all targets, while  J1311 is a clear case, which is similar to AGNs with strong outflows studied by \cite{Karouzos2016}. 

\section{Discussion}
\label{sec:discussion}
Based on the spatially resolved kinematics, we detected the presence of weak outflows in two out of six AGNs in our sample, while
no strong signature of outflows was detected in all six objects based on the single-aperture SDSS spectra. Note that this sample has been 
selected with the same \mbox{[O\,\textsc{iii}]} luminosity limit along with similar stellar mass and host galaxy inclination in comparison with 
AGNs with strong outflows in our previous IFU studies \citep{Karouzos2016, Kang2018}. We discuss the condition and detectability of AGN 
outflows to understand why some luminous AGNs show no strong outflows. 
 
 \subsection{Conditions for the presence of outflows}
\begin{figure}[tbp]
	\begin{center}
		\includegraphics[width=0.45\textwidth,angle=0]{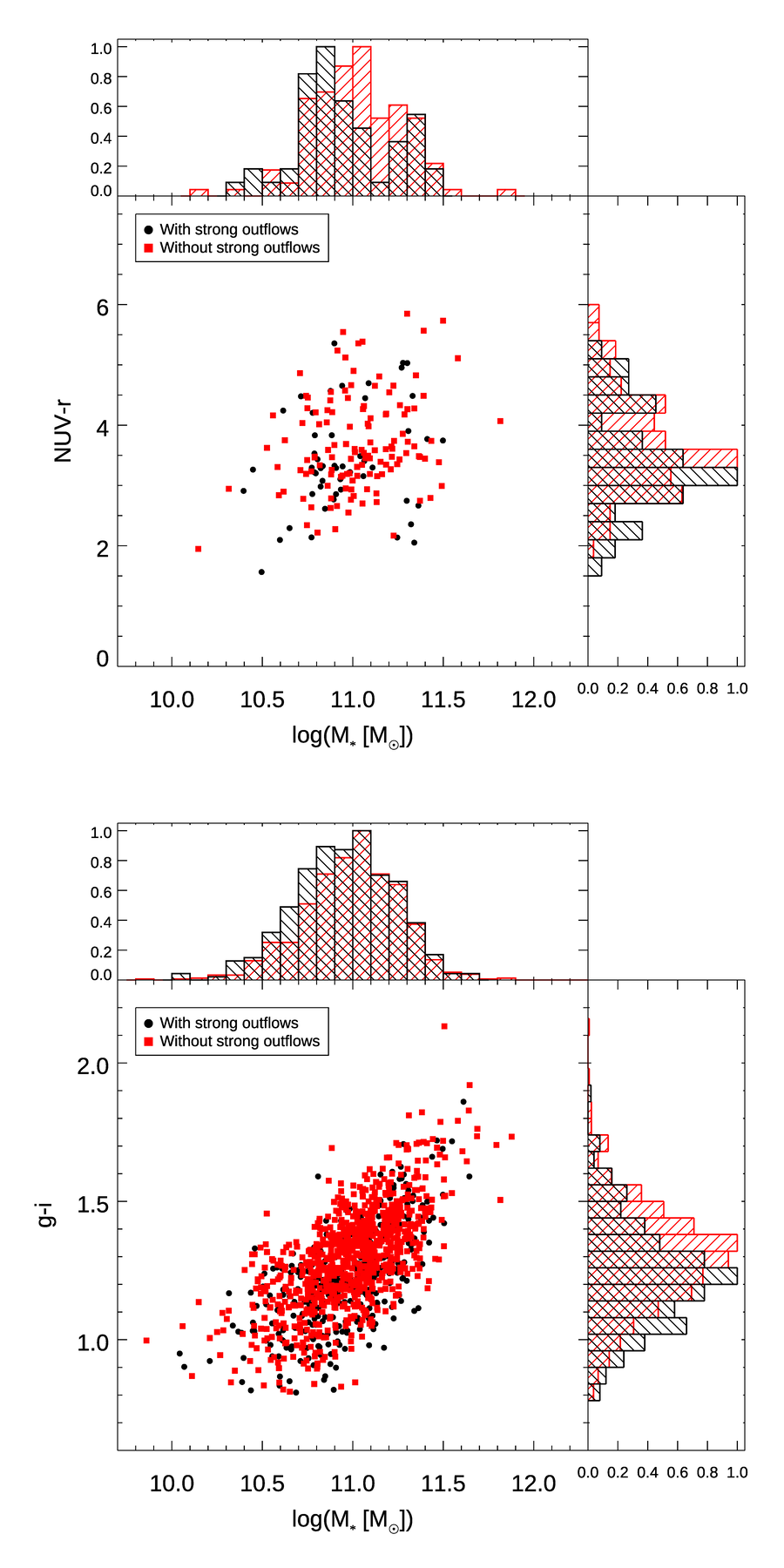}
		\caption{Distribution of $\mbox{NUV}-r$ color (top), $g-i$ color (bottom) and stellar mass for AGNs with and without strong outflows (black and red colors, respectively).}
		\label{fig:gas_fraction_show}
	\end{center}
\end{figure}

One of the key conditions for driving outflows is the gas content in the host galaxies. We investigate the cold gas 
mass fraction using two samples of AGNs, respectively, with and without strong outflows. We use the same sample of the local 
Type 2 AGNs ($\sim$ 39,000 targets at z $<$ 0.3) in \citet{Woo2016}, from which our IFU sample was selected. 
Following selection criteria are applied: (1) We focus on the AGNs with high luminosity, i.e., $L_{\mathrm{\mbox{[O\,\textsc{iii}]};cor}}$ 
($> 10^{42}$ erg s$^{-1}$); (2) To ensure strong outflows, we require the \mbox{[O\,\textsc{iii}]} velocity shift 
$|v_{\mbox{[O\,\textsc{iii}]}}| > 200$ km s$^{-1}$ and \mbox{[O\,\textsc{iii}]} velocity dispersion $\sigma_{\mbox{[O\,\textsc{iii}]}} > 350$ 
km s$^{-1}$; (3) For the AGNs without strong outflows, we limit the \mbox{[O\,\textsc{iii}]} velocity shift $|v_{\mbox{[O\,\textsc{iii}]}}| < 50$ 
km s$^{-1}$ and \mbox{[O\,\textsc{iii}]} velocity dispersion is consistent with stellar velocity dispersion within 20\%. As a result, 
we obtain 383 AGNs with strong outflows and 1051 AGNs without strong outflows, respectively. 

Due to the lack of the multi-wavelength observations to measure gas fraction, we estimate the gas mass fraction by adopting the 
photometric technique (e.g. \citealt{Kannappan2004,Eckert2015}), which
uses a broad-band color as a proxy for the cold gas mass fraction. Since our targets are Type 2 AGNs, the galaxy colors will 
not be significantly affected by the AGN continuum. If any, the effect of AGN on galaxy colors will be similar between two samples. 
Following the calibration by \citet{Eckert2015}, we use the $g-i$ color to derive the HI gas fraction [$M_{HI}/M_{*}$] as
\begin{equation}
log(M_{HI}/M_{*})=-0.984(2.444(g-i)+0.550(b/a))+1.881,
\end{equation}
where $g-i$ is limited between 0.8 and 2.6 magnitude, and $b/a$ is the axis ratio of the galaxy.
As described in \citet{Eckert2015}, the scatter of photometric gas fraction calibration is minimized by including $b/a$ in Equation (2). 
In the case of the molecular gas fraction [$M_{H_{2}}/M_{*}$], we use the $\mbox{NUV}-r$ color and the scaling relation by \citet{Saintonge2011},
\begin{equation}
log(M_{H_{2}}/M_{*})=-0.293(\mbox{NUV}-r-3.5)-1.349.
\end{equation}
Note that these two calibrations have substantial scatter, which introduces relatively large systematic uncertainties in the derived gas fraction. 
Based on the scatter shown in \citet{Eckert2015} and \citet{Saintonge2011}, we estimate that the systematic uncertainty is a factor of 2 for the gas fraction.

In Fig. \ref{fig:gas_fraction_show}, we compare the distributions of $\mbox{NUV}-r$ color and $g-i$ color with the stellar mass distribution for the selected 
AGNs with and without strong outflows. We find that the $\mbox{NUV}-r$ color distribution is not significantly different between the two samples, for given 
the small sample size. In the case of the $g-i$ color, the distribution is somewhat different, showing on average redder $g-i$ color for AGNs without strong 
outflows, while the stellar mass distribution of the two samples is comparable. 

\begin{figure}[tbp]
	\begin{center}
		\includegraphics[width=0.48\textwidth,angle=0]{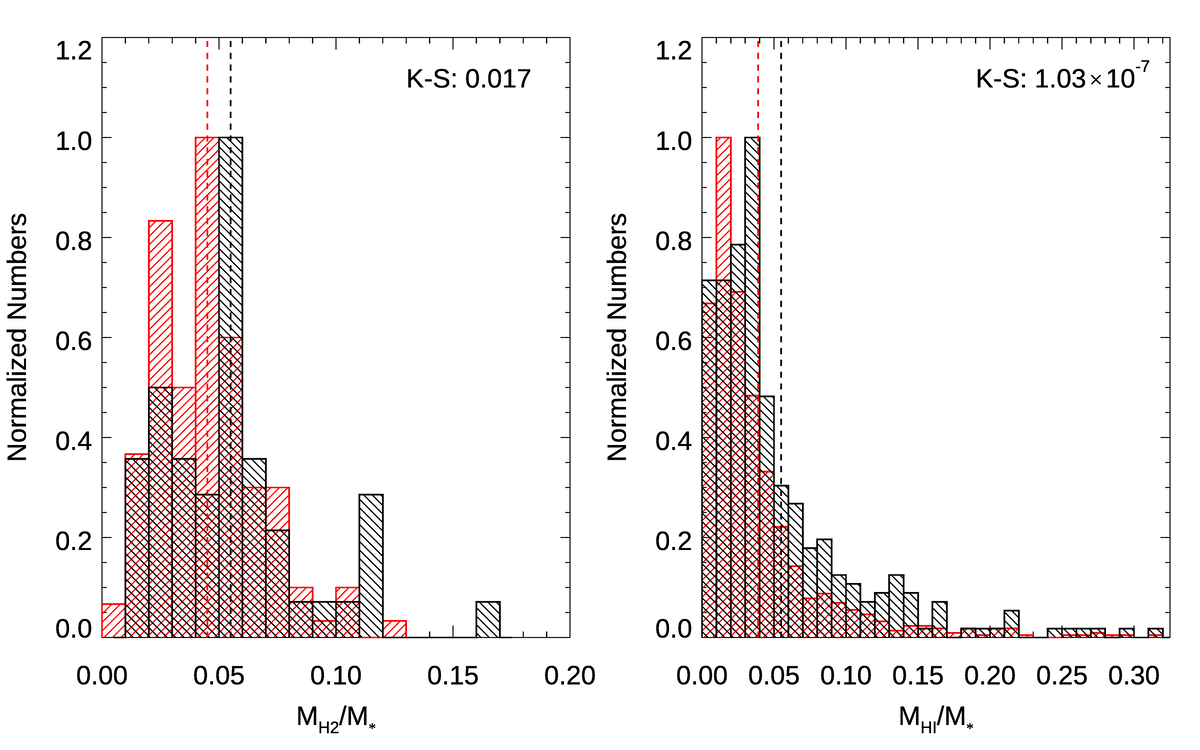}
		\caption{Distribution of molecular gas fraction (left) and the HI gas fraction (right), based on broad-band colors, for AGNs with and 
			without strong outflows (black and red colors, respectively). The mean gas fraction of each sample is denoted with a dotted line, 
			and the p-value of a two-sample K-S test is given at the upper-right corner}
		\label{fig:gas_fraction}
	\end{center}
\end{figure}

In Fig. \ref{fig:gas_fraction}, we present the calculated gas mass fraction using Eq. 1 and 2. 
For the molecular gas fraction, we find no significant difference with the mean fraction of $0.055 \pm 0.031$ and $0.045 \pm 0.022$, 
respectively for AGNs with and without strong outflows. A two-sample Kolmogorov-Smirnov (KS) test also fails to reject the null hypothesis (with 
a probability $p=0.017$) that the two samples are drawn from the same parent distribution. 

In the case of HI gas, the average gas fraction is comparable between the AGNs with and without strong outflows, 
with the mean fraction of $0.055 \pm 0.058$ and $0.039 \pm 0.045$, respectively. However, the KS test rejects the null hypothesis that the 
two samples are drawn from the same parent distribution at a probability value of $p=1.03\times10^{-7}$. This is due to the overall difference 
in the shape of the distribution. This result may suggests that AGNs with and without strong outflows have different distributions of HI gas fraction, 
implying that the detection of outflows is related to the HI gas content. Since the photometric gas fraction technique has large systemic uncertainties 
and should be applied to a large sample for statistical analysis, direct measurements of gas fraction in individual galaxies  based on HI and CO 
observations are required, in order to confirm that HI gas fraction is the key for detecting AGN-driven outflows. Following up studies with 
multi-wavelength data will provide better constrains on the connection between cold gas fraction and AGN outflows.

\subsection{Outflow detectability}

Kpc-scale ionized gas outflows are frequently observed in luminous AGNs, although the outflow fraction varies depending on the definition 
of outflows. Based on a large sample of $\sim$ 39,000 Type 2 AGNs, \citet{Woo2016} adopted the non-gravitational kinematic signature 
(i.e. $\sigma_{\mbox{[O\,\textsc{iii}]}} > \sigma_{\ast}$) to identify gas outflows, finding that the outflow fraction is at least $\sim$50\% 
over the large dynamic range of [OIII] luminosity, while there is a strong increase of outflow fraction as a function of luminosity,
reaching over 80\% at $L_{\mathrm{\mbox{[O\,\textsc{iii}]}}} > 10^{42}$ erg s$^{-1}$. By using the similar analysis method, 
\citet{Rakshit2018} performed a census of ionized gas outflows using a large sample of Type 1 AGNs ($\sim$ 5000 targets at z $<$ 0.3), 
reporting that the outflow fraction of Type 1 AGNs is $\sim$90\%. As Type 1 AGNs have higher luminosity, these two results 
consistently indicate a high outflow fraction in luminous AGNs. Similarly, \citet{Sun2017} set a constrain on the 
outflow fraction of more luminous Type 2 AGNs (i.e., $L_{bol} \gtrsim 10^{46}$ erg s$^{-1}$), reporting a $> 60\%$ and 
possibly $\sim90\%$ occurrence rate. The high occurrence rate of gas outflow indicates that it can persist for a relative long timescale, 
which has been predicted in the theoretical model of AGN-driven outflows \citep{King2011}. As proposed by \citet{Sun2017}, short-term 
AGN variability over a long-term AGN episode with a moderate AGN duty cycle is a possible scenario to explain the above 
high outflow occurrence rate. 

Since the outflow fraction is very high for luminous AGNs, it may be naively expected that for the relatively luminous AGNs in our sample, 
the spatially resolved study could reveal AGN-driven outflows, although we did not find signatures of outflows based on the SDSS spectra.
However, we detect weak outflow signatures only in two targets out of six AGNs. While we find a hint of difference in HI gas fraction between 
AGNs with and without outflows, it remains unclear what determines the outflow occurrence. 

The structure of outflows and the viewing angle can influence the outflow detectability. As described in the 3D biconical 
outflow models of \citet{Bae2016}, several structure parameters (e.g. bicone inclination, dust plane inclination, bicone opening 
angle and dust extinction) can affect the observed gas kinematics (e.g. $\mbox{[O\,\textsc{iii}]}$ velocity and 
velocity dispersion) in the projected plane. If the outflow direction is almost perpendicular to the line-of-sight, then the projected velocity shift 
with respect to the systemic velocity would be minimized, while the velocity dispersion will be relatively small in the line-of-sight if the opening 
angle of the outflows is small \citep{Bae2016}. The intrinsic nature of outflows is yet to be understood except for the outflows in very nearby 
Seyfert galaxies, for which much higher spatial resolution was possible. More detailed comparison of spatially resolved data with the kinematical 
models is required to better understand the detectability of ionized gas outflows.

\section{Summary}
\label{sec:summary}
For a sample of six local (z $<$ 0.1) and luminous (L$_{\mathrm{[OIII]}}>10^{42}$ erg s$^{-1}$) Type 2 AGNs, which were
selected as AGNs without strong signature of outflows from a large sample of $\sim$39,000 Type 2 AGNs, 
we performed Gemini/GMOS-IFU observations to investigate the spatially resolved kinematics
and the presence of outflows. We summarize the main results as below.

\begin{itemize}
\item Ionized gas kinematics in our sample are governed by various physical processes, 
including merging, AGN-driven outflows and host galaxy gravitational potential. 
In two targets (i.e., J1019 and J1058), we find significant difference of the kinematics between ionized gas and stars, which can be explained by AGN-driven outflows. 

\item Based on the spatially resolved kinematics, we find kinematic signatures of outflows for two AGNs (i.e., J1019 and J1058), 
while the integrated SDSS spectra show no significant outflow signature. The VVD diagram of these two ANGs is consistent with the prediction of outflow kinematics. 
However, velocity and velocity dispersion of outflows are relatively low, suggesting weak outflows. These results suggest that the outflow fraction derived using integrated 
spectra can be underestimated, highlighting the importance of spatially-resolved observation in identifying signatures of AGN-driven outflows. 

\item While the central part is dominated by AGN photoionization, we find a signature of mixing from AGN and star-formation at the edge of the emission region in 
four AGNs (J0843, J1019, J1058, J1156). One AGN (J1311) clearly shows a ring-like structure of star forming region.

\item Based on the indirect estimates of HI gas fraction estimated with the $g-i$ color, we find a hint of the difference in
the HI gas fraction between AGNs with and without strong outflows. This result implies that the presence of outflows may depend on the gas fraction.
\end{itemize}

\acknowledgments{ We thank the anonymous referee for useful comments, which improved the clarity of the manuscript. 
	This research was supported by the National Research Foundation of Korea (NRF) grant funded by the Korea government (MEST) (No.  2016R1A2B3011457). 
	This work was supported by K-GMT Science Program (PID: GN-2016A-Q-19) funded through the Project BIG3: "From Big Bang to Big Data with Big Eyes" operated by Korea 
	Astronomy and Space Science Institute (KASI). Based on observations obtained at the Gemini Observatory processed using the Gemini IRAF package, which is operated by 
	the Association of Universities for Research in Astronomy, Inc. (AURA) under a cooperative agreement with the NSF on behalf of the Gemini partnership: the US National 
	Science Foundation (NSF), the Canadian National Research Council (NRC), the Chilean Comisi\'{o}n Nacional de Investigaci\'{o}n Cientifica y Tecnol\'{o}gica (CONICYT), the 
	Brazilian Minist\'{e}rio da Ci\^{e}ncia, the Argentinean Ministerio de Ciencia, Tecnolog\'{i}a e Innovaci\'{o}n Productiva, Tecnologia e Inova\c{c}\~{a}o and the Korea Astronomy and Space Institute (KASI).}

\begin{appendix}
\label{sec:appendix}
In this section, we provide the detailed comments for individual targets.
\vspace{10pt}
\section{J084344+354942}
\label{sec:J0843}
At z = 0.0541, the GMOS FOV of J0843 covers a $3.6\times5.4$ kpc region with a spatial resolution of 0.7 kpc. 
J0843 appears as a merging galaxy with tidal features. \mbox{[O\,\textsc{iii}]} emission 
is detected (with S/N $>$ 3) in the central $3.6\times2.6$ kpc region, while H$\alpha$ emission is more extended. 
The stellar velocity map shows a velocity gradient 
in the NW-SE direction. The ionized gas velocity map is complex, probably disturbed by the merging process. 
As shown in the BPT classification map, AGN photoionization is dominated in the area with significant detection 
of emission lines, while there are signatures of LINER emission and star-formation at the edge.

\section{J101936+193313}
\label{sec:J1019}
At z = 0.0647, the GMOS FOV of J1019 covers a $4.3\times6.4$ kpc region with a spatial resolution of $0.83$ kpc.
With the axis ratio b/a=0.60, J1019 appears as a slightly inclined disk galaxy elongated in the N-S direction. 
\mbox{[O\,\textsc{iii}]} and H$\alpha$ emission present a similar spatial distribution with a symmetric shape
along the NE-SW direction. 
The stellar velocity map reveals a clear rotation pattern with the kinematic major axis along the NE-SW direction. 
The gas velocity map shows an opposite trend compared to that of stars, possibly suggesting a counter-rotation. 
The bi-conical outflows driven by AGNs is consistent with the observed gas kinematics, if the outflow is along the kinematic 
major axis. As shown in the BPT classification map, AGN photoionization is dominant, while there are signatures of LINER 
emission and star-formation at the edge of the emission line region.

\section{J105833+461604}
\label{sec:J1058}
At z = 0.0397, the GMOS FOV of J1058 covers a $2.7\times4.0$ kpc region with a spatial resolution of 0.53 kpc. 
J1058 is a face-on disk galaxy with a b/a=0.89 along with a clear bar structure. H$\alpha$ 
emission (with S/N $>$ 3) is distributed in the central $2.7\times2.9$ kpc region, which is more extended than the 
distribution of \mbox{[O\,\textsc{iii}]} emission. The stellar velocity 
map reveals a rotation pattern. The kinematic major axis of the ionized gas is misaligned with respect 
to that of the stars by 40 degree, which could be due to the bi-conical outflows driven by AGN. The BPT classification 
map shows AGN dominated photoionization, while LINER emission is detected at the edge of the emission line region.

\section{J115657+550821}
\label{sec:J1156}
At z = 0.0800, the GMOS FOV of J1156 covers a $5.3\times8.0$ kpc region with a spatial resolution of
$1.6$ kpc. J1156 is a face-on disk galaxy with b/a= 0.84. \mbox{[O\,\textsc{iii}]} and H$\alpha$ 
emission is concentrated within the $4.0$ kpc region. 
The stellar velocity map shows no clear rotation pattern while the gas velocity map shows positive (redshifted) velocities, 
which may be signatures of outflows, but the velocity and velocity dispersion of the ionized gas are relatively small.
AGN photoionization is dominant at the center, while there are signatures of LINER emission and star-formation at 
the edge of the emission line region.

\section{J131153+053138}
\label{sec:J1311}
At z = 0.0873, the GMOS FOV of J1311 covers a $5.8\times8.7$ kpc region with a spatial resolution of
$1.1$ kpc. With b/a= 0.85, J1311 appears as a face-on disk galaxy with an outer ring 
structure. \mbox{[O\,\textsc{iii}]} emission is centrally concentrated in the 4 kpc region. 
In contrast, H$\alpha$ emission is more extended, including a bright 
central component and a weak flux region, that is reaching out to the edge of the FOV. The stellar velocity map
reveals a rotation pattern with the kinematic major axis along the E-W direction.
The H$\alpha$ velocity map is similar to the stellar velocity map, while the \mbox{[O\,\textsc{iii}]} velocity 
map is dominated by positive (redshifted) velocities, which could be due to the AGN-driven outflows. However, \OIII\
velocity is relatively small, particularly at the center. 
As shown in the BPT classification map, the AGN photoionization is dominated in the central region, 
while there is a ring-like structure of star-forming region at the edge.

\section{J161756+221943}
\label{sec:J1617}
At z = 0.1020, the GMOS FOV of J1617 covers a $6.8\times10.3$ kpc region with a spatial resolution of
$2$ kpc. With b/a= 0.85, J1617 appears as a face-on disk galaxy. \mbox{[O\,\textsc{iii}]} 
emission is concentrated within the central 2 kpc region, while H$\alpha$ emission is more extended. 
The stellar velocity map reveals a rotation pattern with the kinematic major axis along the NW-SE direction. 
The gas velocity map is similar to the stellar velocity map, showing a similar rotation pattern. 
The BPT classification map shows AGN dominated photoionization. 

\end{appendix}

\bibliographystyle{aasjournal}
\bibliography{GMOS_IFU_16A_astroph}

\end{document}